\documentclass{article}

\usepackage{PRIMEarxiv}
 
\usepackage{amsmath,amssymb,amsfonts}
\usepackage{array}
\usepackage{url}
\usepackage{orcidlink}
\usepackage{tabularx}
\usepackage{multirow}
\usepackage{booktabs}    
\usepackage{makecell}   

\usepackage[utf8]{inputenc} 
\usepackage[T1]{fontenc}    
\usepackage{hyperref}       
\usepackage{url}            
\usepackage{booktabs}       
\usepackage{amsfonts}       
\usepackage{nicefrac}       
\usepackage{microtype}      
\usepackage{lipsum}
\usepackage{fancyhdr}       
\usepackage{graphicx}       
\graphicspath{{media/}}     
\usepackage{authblk}
\title{Hypergraph Multi-Modal Learning for EEG-based Emotion Recognition in Conversation
\thanks{\textit{\underline{Citation}}: 
\textbf{Z Kang, Y Li, S Gong, W Zeng, H Yan, L Bian, Z Zhang, WT Siok, N Wang. Hypergraph Multi-Modal Learning for EEG-based Emotion Recognition in Conversation. Neural Networks, 2026, 109057, DOI:10.1016/j.neunet.2026.109057.}} 
}

\author[1,$\dagger$]{Zijian Kang}
\author[1,2,$\dagger$]{Yueyang Li}
\author[1]{Shengyu Gong}
\author[1,$\star$]{Weiming Zeng}
\author[3]{Hongjie Yan}
\author[2,4]{Lingbin Bian}
\author[5]{Zhiguo Zhang}
\author[2]{Wai Ting Siok}
\author[2,$\star$]{Nizhuan Wang}

\affil[1]{\textit{Lab of Digital Image and Intelligent Computation, Shanghai Maritime University}, Shanghai 201306, China}
\affil[2]{\textit{Department of Chinese and Bilingual Studies, The Hong Kong Polytechnic University}, Hong Kong SAR, China}
\affil[3]{\textit{Affiliated Lianyungang Hospital of Xuzhou Medical University}, Lianyungang 222002, China}
\affil[4]{\textit{The State Key Laboratory of Brain and Cognitive Sciences, The University of Hong Kong}, Hong Kong SAR, China}
\affil[5]{\textit{The Institute of Computing and Intelligence, Harbin Institute of Technology Shenzhen}, Shenzhen 518000, China}
\affil[$\dagger$]{Co-first authors}
\affil[$\star$]{Correspondence: wangnizhuan1120@gmail.com; zengwm86@163.com}

\begin{document}
\maketitle

\begin{abstract}

Emotion Recognition in Conversation (ERC) is valuable for diagnosing health conditions such as autism and depression, and for understanding the emotions of individuals who struggle to express their feelings. Current ERC methods primarily rely on semantic, audio and video data but face significant challenges in integrating physiological signals such as Electroencephalography (EEG), which has low signal-to-noise ratios, inter-subject variability, and temporal alignment issues. This research proposes Hypergraph Multi-Modal Learning (Hyper-MML), a novel framework for identifying emotions in conversation. Hyper-MML effectively integrates EEG with audio and video information to capture complex emotional dynamics. Firstly, we introduce an Adaptive Brain Encoder with Mutual-cross Attention (ABEMA) module for processing EEG signals. This module captures emotion-relevant features across different frequency bands and adapts to subject-specific variations through hierarchical mutual-cross attention mechanisms. Secondly, we propose an Adaptive Hypergraph Fusion Module (AHFM) to actively model the higher-order relationships among multi-modal signals in ERC. Experimental results on the EAV and AFFEC datasets demonstrate that our Hyper-MML model significantly outperforms current state-of-the-art methods. The proposed Hyper-MML can serve as an effective communication tool for healthcare professionals, enabling better engagement with patients who have difficulty expressing their emotions. The official implementation codes are available at https://github.com/NZWANG/Hyper-MML.

\end{abstract}


\section{Introduction}\label{Introduction}

\subsection{Emotion Recognition in Conversation (ERC)}

\begin{figure}
    \centering
    \includegraphics[width=1\linewidth]{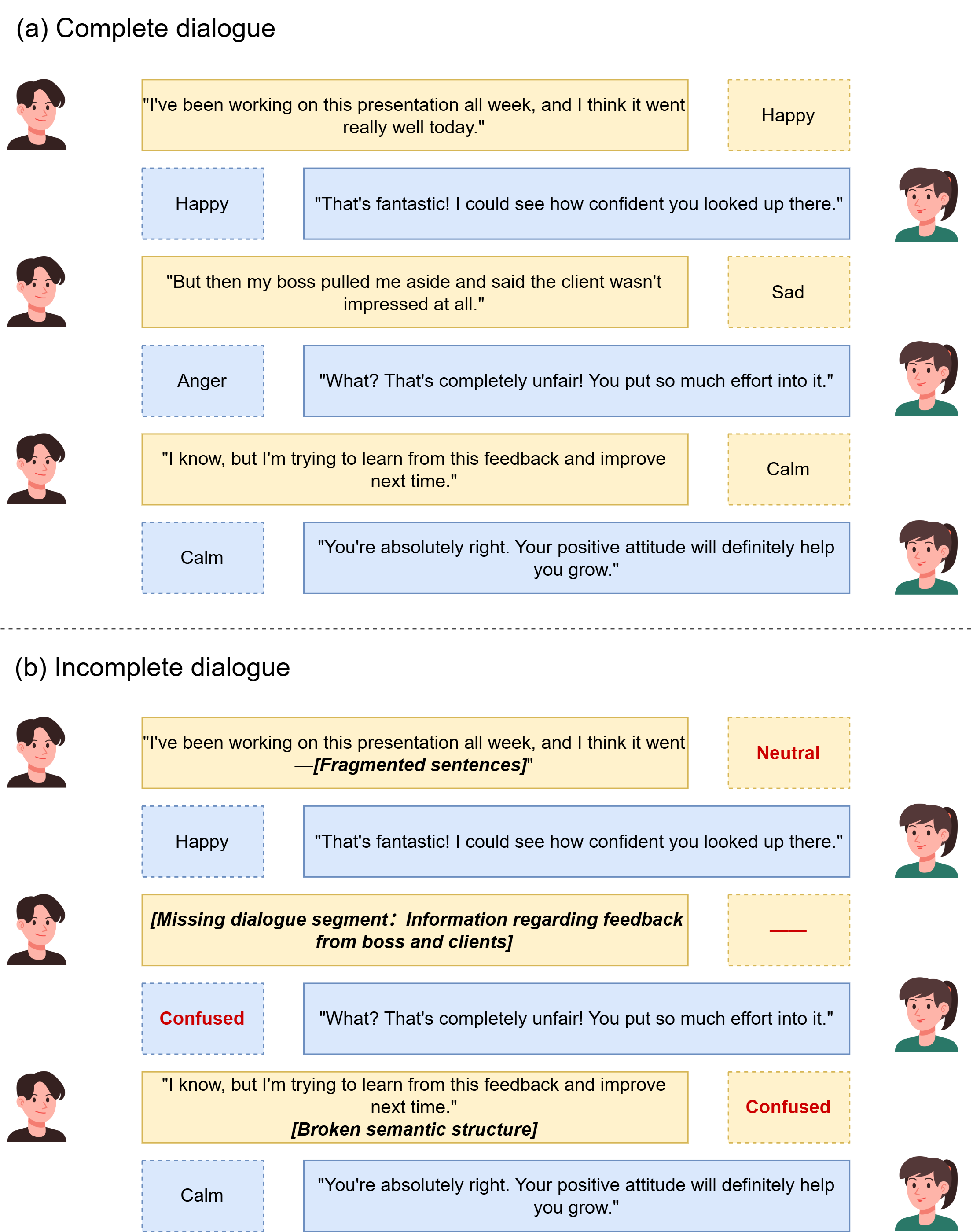}
    \caption{Illustration of emotion recognition in conversational text. The complete dialogue (a) shows coherent emotional transitions and logical semantic structure. Incomplete dialogue (b), such as fragmented sentences, missing dialogue segments and broken semantic structure, causes emotion label ambiguity within a conversation. This leads to emotion misclassification and contextual confusion.}
    \label{fig:incomplete_text}
\end{figure}

Emotion Recognition in Conversation (ERC) holds significant potential for assessing mental conditions such as autism and depression. Recent studies suggest that individuals with these conditions frequently exhibit unique communication challenges, including speech impairments, emotional disturbances, literal interpretation of questions, and difficulty sustaining coherent dialogue \cite{hervas2023autism}. Current ERC research primarily focuses on transcribed text from spoken dialogue, supplemented by visual (e.g., facial expressions) and acoustic (e.g., intonation, loudness) cues \cite{poria2019emotion}. These methods typically analyze complete, uninterrupted dialogue transcripts, integrating multi-modal data to contextualize individual utterances. Context modeling in ERC generally incorporates three key elements: 1) the content of previous exchanges, 2) the timing of conversational turns, and 3) speaker-specific details such as identity and evolving emotional states \cite{hu2021mmgcn}. However, the semantic structure of dialogue is vulnerable to disruptions such as fragmented sentences and missing segments, which distort the logical relationships between utterances. This resulting incoherence severely degrades the performance of emotion recognition models, ultimately constraining their practical applications in clinical settings for disorders like autism and depression. Figure \ref{fig:incomplete_text} provides a demonstration of this phenomenon, showing how such fragmentation leads to significant emotion misclassification. To address these limitations, psychotherapists require robust physiological indicators that remain reliable even with incomplete conversational data. 

Physiological signals -- particularly electroencephalography (EEG) data -- provide a direct window into neural activity and emotional states, surpassing text-based methods in objectivity and immediacy \cite{li2022eeg,li2025tale,lujan2024two}. While textual analysis depends on extended linguistic context, EEG signals operate on shorter timescales, making them ideal for detecting transient emotional shifts (e.g., sudden frustration or momentary joy) in real time. Thus, by integrating EEG signals with text-based modalities, clinicians can address key limitations of language-driven approaches, such as distortions caused by fragmented or incomplete dialogue. EEG's  intrinsic nature avoids language-related biases, enabling clearer and more objective emotion measurement. Furthermore, combining EEG with multi-modal data (e.g., audio and video) outperforms single-source EEG analysis, enhancing diagnostic accuracy \cite{zhao2024feature}. This integrative framework will allow psychologists or clinicians to correlate physiological responses (e.g., brainwave patterns) with behavioral cues (e.g., voice tone, facial expressions), constructing a comprehensive emotional profile, which supports the development of tailored treatment strategies that better address individual patient needs \cite{chen2025eeg}. 

Despite these advantages, integrating EEG signals into multi-modal ERC frameworks presents technical challenges that have constrained their widespread adoption. EEG signals inherently suffer from low signal-to-noise ratios, which are further compromised by unavoidable Electromyography (EMG) artifacts during natural conversation, including muscle movements from speaking, facial expressions, and head motion that can significantly contaminate the neural signal. Moreover, the inter-subject variability in brain responses poses a critical challenge, as individuals exhibit distinct neural activation patterns for identical emotional states due to differences in brain anatomy, electrode placement sensitivity, and underlying neurophysiological characteristics. In addition, the temporal dynamics of emotional states and cross-modal synchronization issues further complicate the integration process, requiring sophisticated architectures capable of handling complex spatio-temporal dependencies across physiological and behavioral modalities.

In multi-modal ERC tasks,  graph neural networks (GNNs) are often employed to model interactions by capturing contextual and multi-modal data (e.g., text, audio, video). However, GNNs face a critical limitation: they can only model complex interactions by chaining together simple pairwise relationships (e.g., between two nodes at a time). This sequential approximation of high-order relationships -- such as group dynamics or multi-modal dependencies -- often leads to suboptimal accuracy. Hypergraph theory overcomes this limitation by supporting high-order connections (e.g., linking three or more nodes simultaneously), enabling direct modeling of intricate multi-modal interactions. For instance, a hyperedge could connect a speaker's utterance, their facial expression, and a listener's reaction in a single interaction step. This capability makes hypergraphs a more precise and efficient framework for multi-modal ERC tasks \cite{yadati2019hypergcn}.

\subsection{Our Contribution}

We propose a novel Hypergraph Multi-Modal Learning (Hyper-MML) framework for ERC that achieves state-of-the-art (SOTA) performance on the EAV \cite{lee2024eav} and AFFEC \cite{sekiavandi2025advancing} datasets. Our model advances ERC in following three key innovations:
\begin{enumerate}
\renewcommand{\labelenumi}{\arabic{enumi})}
    \item \textbf{Hypergraph Multi-Modal Learning Framework (Hyper-MML)}:
We introduce an end-to-end architecture that integrates EEG signals with audio and video data to model complex emotional dynamics in conversations. Unlike traditional language-centric approaches, Hyper-MML directly leverages physiological (EEG) and behavioral (audio-video) cues, bypassing the limitations of language-based ambiguity or incomplete dialogue transcripts while addressing the temporal alignment challenges through unified time-window segmentation.

    \item \textbf{Adaptive Brain Encoder with Mutual-cross Attention(ABEMA)}:
A specialized EEG encoder captures emotion-relevant patterns across frequency bands while adapting to individual brain characteristics. ABEMA's hierarchical attention mechanism models both within-band and cross-band relationships, enabling robust feature extraction from noisy EEG signals, while EEG preprocessing and frequency-domain processing mitigate noise and EMG artifact contamination.

    \item \textbf{Adaptive Hypergraph Fusion Module (AHFM)}:
A specialized fusion module enhances cross-modal interaction within hypergraph structures. This module employs adaptive weighted aggregation to dynamically prioritize the most informative modalities (e.g., emphasizing EEG during subtle emotional shifts or audio during tone-based cues) and models long-range emotional patterns through inter-segment hyperedges. This strategy optimizes information propagation across modalities, significantly improving emotion recognition accuracy.
\end{enumerate}

The remainder of this paper is organized as follows: Section \ref{Related work} reviews prior work in multi-modal emotion recognition and hypergraph learning. Section \ref{Methodology} introduces the architecture of our proposed Hyper-MML model. Section \ref{Experiments} describes the datasets and experimental setup. Results and discussion are presented in Section \ref{Results and Discussion}. Finally, Section \ref{Conclusion and future work} concludes the paper and outlines future research directions.

\section{Related work}\label{Related work}

\subsection{Multi-modal Emotion Recognition}

ERC has become essential for healthcare diagnostics, particularly in autism and depression assessment \cite{poria2019emotion}. This field is evolving from early text-centric approaches to sophisticated multi-modal frameworks that integrate textual, acoustic, and visual information. Early approaches focused on textual analysis, incorporating conversational history, temporal dynamics, and speaker information \cite{majumder2019dialoguernn}. Recent advances have introduced graph-based approaches such as MMGCN \cite{hu2021mmgcn}, which model complex conversational relationships through graph neural networks. However, these text-dependent methods struggle with fragmented dialogue data, common in clinical settings with communication-impaired patients. Cross-modal reconstruction techniques address missing modalities \cite{liu2024contrastive}, but suffer from modality gap issues \cite{liang2022mind}, where different modalities embed at distinct distances, affecting reconstruction accuracy. This limitation motivates the exploration of alternative modalities that can provide robust emotional indicators independent of textual coherence.

EEG provides direct access to neural activity and emotional states \cite{gu2026eeg, li2026sc}, offering advantages in objectivity, temporal resolution, and language independence compared to behavioral modalities \cite{li2022eeg, wang2024neuroimaging}. EEG-based emotion recognition has progressed from traditional frequency analysis using Power Spectral Density (PSD) and Differential Entropy (DE) features to deep learning approaches including CNNs \cite{lawhern2018eegnet}, graph methods \cite{jin2024pgcn,li2024efficient}, and temporal modeling \cite{ma2019emotion,zhang2018spatial}. Recent work captures spatial asymmetry and temporal dynamics \cite{ding2022tsception} while modeling brain connectivity through graph networks \cite{tian2025accnet}. However, existing EEG emotion recognition primarily uses passive stimuli (images, videos, music) rather than natural conversational interactions. This gap is significant because the conversational emotions exhibit different temporal evolution and contextual dependencies compared to stimulus-induced emotions. 

Fortunately, recent public datasets help to address this limitation. The EAV dataset \cite{lee2024eav} provides the first public multi-modal EEG-audio-video resource for conversational emotion recognition, featuring 8,400 interactions from 42 participants across five emotional states. The AFFEC dataset \cite{sekiavandi2025advancing} includes EEG, eye-tracking, galvanic skin response (GSR), and facial video data from 73 participants, distinguishing felt versus perceived emotions in face-to-face interactions. Initial research demonstrates multi-modal integration feasibility with conventional modalities \cite{yin2025eeg}, though sophisticated fusion mechanisms for multi-modal emotional interactions remain underdeveloped.

\subsection{Hypergraph Learning}

\begin{figure*}
    \centering
    \includegraphics[width=0.8\linewidth]{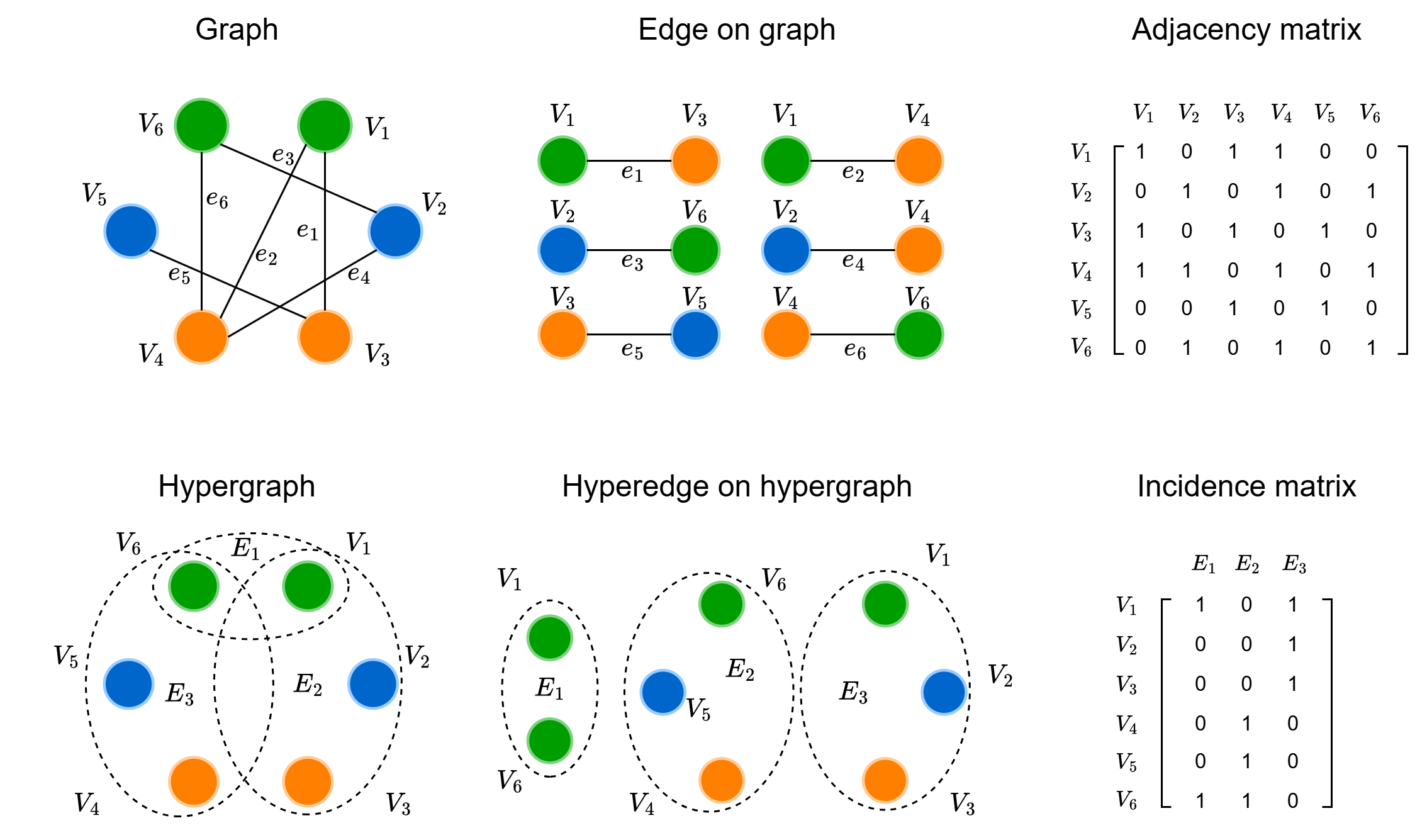}
    \caption{Representation of Graphs and Hypergraphs. The upper left section displays a graph and its edges, while the upper right section presents the adjacency matrix, which describes the connections between nodes. The lower section features a hypergraph and its hyperedges, with the right side showing the incidence matrix, which indicates the relationships between hyperedges and nodes. }
    \label{fig:hypergraph}
\end{figure*}

As illustrated in Figure \ref{fig:hypergraph}, conventional graphs utilize adjacency matrices to encode binary or weighted relationships, limiting their expressive power to simple pairwise interactions. However, many real-world scenarios involve complex higher-order relationships that cannot be adequately captured by pairwise connections. Hypergraphs extend edges to hyperedges, connecting multiple nodes simultaneously to model group interactions and multi-way dependencies \cite{antelmi2023survey}. Hypergraph learning employs incidence matrices instead of adjacency matrices, providing enhanced flexibility for complex relational structures \cite{yadati2019hypergcn,chitra2019random}. This transition of higher-order relationship modeling in hypergraph offers enhanced capabilities for sophisticated relational reasoning.

In the context of EEG-based multi-modal ERC, hypergraphs offer several compelling advantages over traditional graph-based approaches. The simultaneous interaction among EEG signals, audio features, and video cues during emotional expression represents a natural higher-order relationship that cannot be effectively decomposed into pairwise connections without losing critical information. Recent applications of hypergraph learning in multi-modal tasks have demonstrated superior performance in capturing complex correlations \cite{gao2022hgnn+}, with successful implementations in diverse domains including recommender systems \cite{xia2022self}, sleep stage classification \cite{liu2024exploiting}, and drug-target interaction prediction \cite{ruan2021exploring}. In terms of ERC, hypergraphs enable the direct modeling of multi-way dependencies between different modalities within the same temporal context, as well as cross-temporal relationships that span multiple conversation turns. The recent emergence of hypergraph-based approaches in ERC \cite{yi2024multimodal} has shown promising results, though this method primarily focuses on traditional modalities and has not fully explored the integration of physiological signals like EEG. Furthermore, existing approaches often employ fixed hypergraph structures, limiting their adaptability to the dynamic nature of emotional states and the varying importance of different modalities across different emotional contexts, highlighting the need for adaptive hypergraph learning mechanisms that can dynamically adjust to the complex interplay of multi-modal emotional cues in conversational settings.

\section{Methodology}\label{Methodology}

\subsection{Problem Formulation of Multi-modal ERC}

Multi-modal ERC aims to identify participants' emotional states from multi-modal interactions. Generally, we segment each utterance $u_i$ in a conversation sequence $u_i (i = 1, \dots, N)$ into segments based on fixed time intervals $\Delta t$: $u_i = \{s_{i1}, s_{i2}, ..., s_{i{k_i}}\}$. Each segment $s_{ij}$ contains three modalities: EEG signals $E_{ij} \in \mathbb{R}^{C \times L}$, where $C$ is the number of channels and $L$ is the length of time window), audio features $A_{ij} \in \mathbb{R}^{d_a}$, and video features $V_{ij} \in \mathbb{R}^{d_v}$. Our objective is to learn a mapping function $f: (E_{ij}, A_{ij}, V_{ij}) \rightarrow y_i$, where $y_i \in Y$ represents the emotion category.

\begin{figure*}[t]
	\centering
	\includegraphics[width = 1\textwidth]{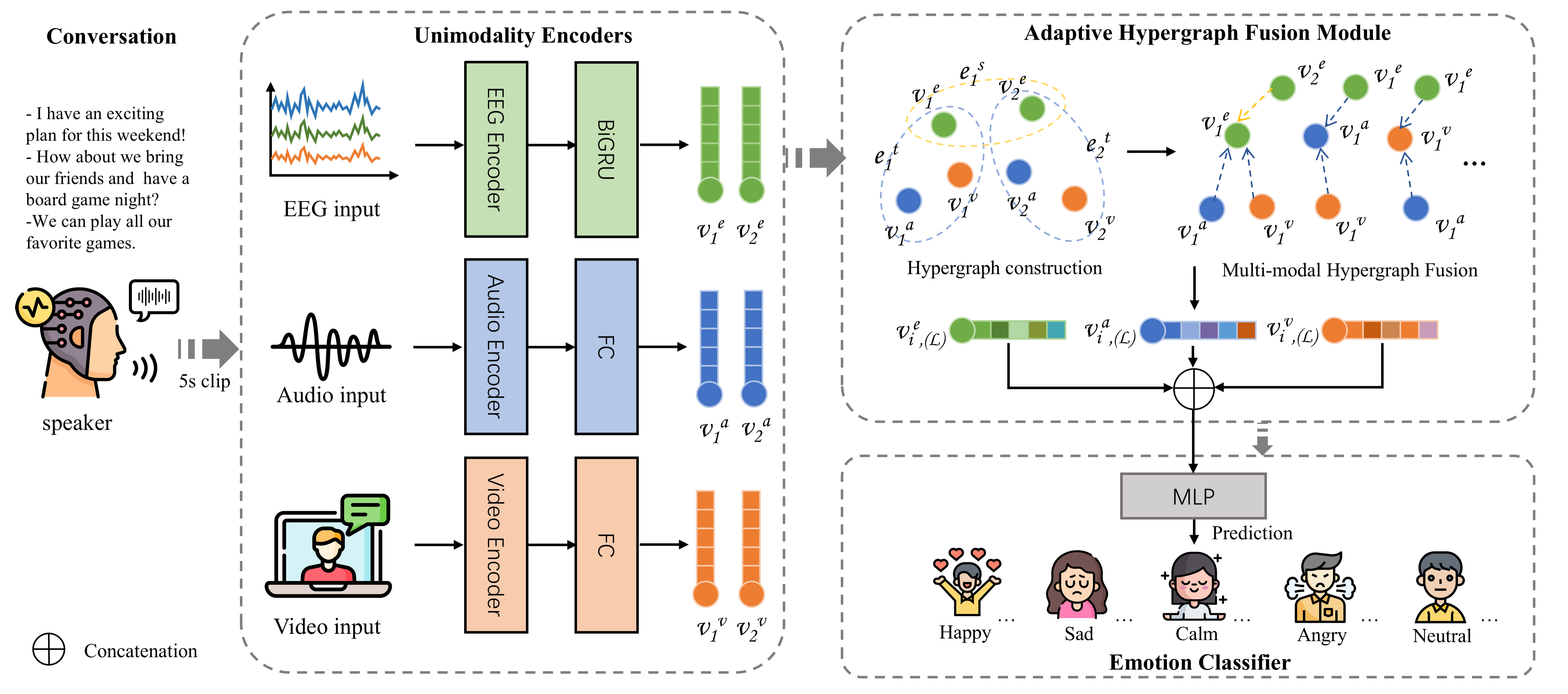}
	\caption{Overall framework of the Hyper-MML.}
	\label{framework}
\end{figure*}

As illustrated in Figure \ref{framework}, we present the Hyper-MML framework to achieve this goal, which effectively models higher-order relationships among multi-modal signals through hypergraph structures, with particular emphasis on fusing EEG signals with traditional modalities to achieve accurate emotion recognition in conversation segments.

\subsection{ABEMA for EEG Embedding}

\begin{figure*}[t]
    \centering
    \includegraphics[width=1\linewidth]{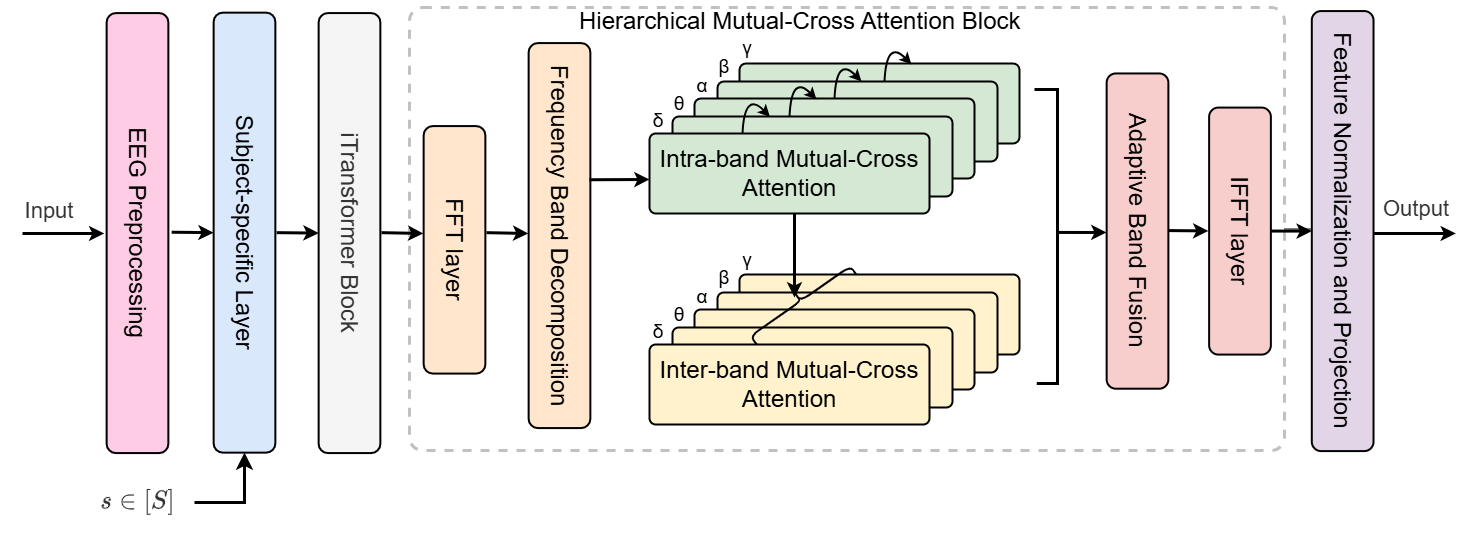}
    \caption{ABEMA for EEG embedding}
    \label{fig:ABEMA}
\end{figure*}

EEG directly reflects neural activity and emotional states, but effectively encoding them for emotion recognition remains challenging due to high dimensionality, low signal-to-noise ratio, and subject-specific variations. To address these challenges, we propose the Adaptive Brain Encoder with Mutual-cross Attention (ABEMA) of EEG (Figure \ref{fig:ABEMA}), which captures emotion-relevant features across different frequency bands while adapting to individual differences. ABEMA transforms raw EEG signals into compact, information-rich embeddings optimized for multi-modal feature fusion. 

\begin{figure*}[t]
    \centering
    \includegraphics[width=1\linewidth]{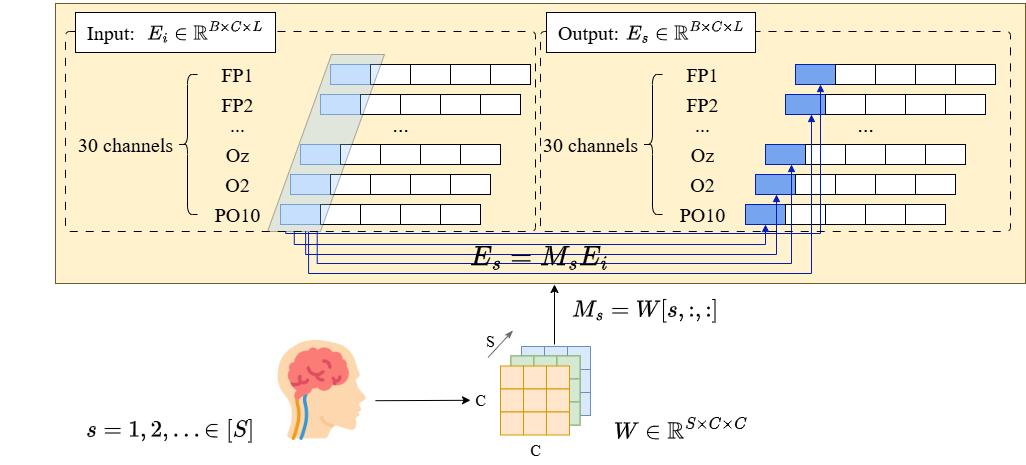}
    \caption{The transformation process of Subject-specific Layer.}
    \label{fig:subjectLayer}
\end{figure*}  

\subsubsection{EEG Preprocessing and Artifact Mitigation}

Our approach builds upon established methodologies from leading multi-modal emotion datasets. The EAV dataset\cite{lee2024eav} demonstrated effective signal quality maintenance through controlled experimental environments with standardized lighting, professional BrainAmp systems with impedance monitoring below 10$k\Omega$, and strategic preprocessing including high-pass filtering (>0.5Hz) and band-pass filtering (50Hz) to mitigate facial electromyographic (EMG) and electrical line noise. Similarly, the AFFEC dataset\cite{sekiavandi2025advancing} validated that even minimal preprocessing approaches—using controlled temperature and lighting conditions with g.tec hiamp 64-channel systems—can achieve robust performance well above chance levels, confirming that foundational signal quality measures are sufficient for natural conversational contexts. These precedents inform our comprehensive artifact management strategy, while adapting these approaches to our specific experimental requirements.

To address EMG artifacts inevitable during natural conversation, we implement a comprehensive preprocessing pipeline tailored to our specific experimental conditions. Raw EEG signals undergo band-pass filtering (0.5-50Hz) using a fourth-order Butterworth filter to remove baseline drift and high-frequency noise. We then apply Independent Component Analysis (ICA) using the extended Infomax algorithm to identify and remove EMG-related components based on their spectral characteristics (high power in frequencies >30Hz) and spatial distribution patterns, effectively eliminating major muscular artifacts while preserving emotion-related neural components in lower frequency ranges.

\subsubsection{Subject-specific Layer} 

EEG signals exhibit significant variations across different subjects due to differences in brain structures, electrode placement, and neurophysiological characteristics. This inter-subject variability reduces model generalization capability. To address this issue, we adopt the subject-specific layer from previous work \cite{li2024neural,defossez2023decoding}. This layer serves as a calibration step to normalize individual differences in electrode positioning, impedance variations, and anatomical differences, which primarily manifest as linear scaling and mixing effects \cite{li2023cross}.

\textbf{Architecture Design:} As shown in Figure \ref{fig:subjectLayer}, the subject-specific layer maintains a 3D parameter tensor $\mathbf{W} \in \mathbb{R}^{S \times C \times C}$, where each slice $\mathbf{W}_s \in \mathbb{R}^{C \times C}$ represents a learnable subject-specific transformation matrix for subject $s$ $\in [1,S]$, $S$ is the number of subjects, and $C$ is the number of EEG channels. Subject identifiers are encoded as integer indices  obtained from dataset metadata and passed alongside EEG data during processing.

\textbf{Dynamic Transformation:} Given an input EEG signal $E_i \in \mathbb{R}^{B \times C \times L}$ from subject $s$, we dynamically select the corresponding transformation matrix $\mathbf{M}_s = \mathbf{W}[s,:,:]$ and apply  channel-wise linear transformation:
\begin{equation}
    \begin{aligned}
    E_s[b, :, t] &= \text{SubjectLayer}(E_i[b, :, t], s) \\
    &= \mathbf{M}_s \cdot E_i[b, :, t] \quad \forall b \in [1,B], t \in [1,L]
    \end{aligned}
\end{equation}
where the Einstein summation performs matrix multiplication along the channel dimension for each batch $b$, time step $t$, effectively applying the subject-specific linear transformation $\mathbf{M}_s$ to each temporal slice of the EEG signal. 
Each output channel is the weighted sum of all input channels. This performs a linear transformation of the channel activations at each time step, effectively recalibrating the spatial patterns to account for individual differences.

\textbf{Implementation and Optimization:} The transformation is implemented as a 1×1 convolution without bias or activation function \cite{defossez2023decoding}, applied along the channel dimension. Each subject-specific matrix is initialized as an identity matrix $\mathbf{M}_s^{(0)}$ with small Gaussian noise ($\sigma = 0.01$)
\begin{equation}
    \mathbf{M}_s^{(0)} = \mathbf{I} + \epsilon \mathbf{N}(0, \sigma^2)
\end{equation}
to preserve signal characteristics during early training. During training, gradients are computed only when samples from the corresponding subject are present in the current batch, enabling efficient parameter updates for individual differences. The subsequent hierarchical mutual-cross attention mechanisms and frequency-domain processing in our ABEMA module are specifically designed to capture the nonlinear temporal and spectral patterns crucial for emotion recognition.

The resulting transformed EEG signals $E_s$ $\in \mathbb{R}^{B \times C \times L}$ retain the original temporal and spatial structure while being adapted to account for subject-specific characteristics, providing a robust foundation for subsequent processing.

\subsubsection{iTransformer Block}
EEG signals from different brain regions exhibit complex spatial correlations that are crucial for emotion recognition. To effectively capture these inter-channel relationships while preserving temporal structure,  we employ an improved Transformer architecture \cite{liu2023itransformer} for spatial feature extraction:
\begin{equation} 
	E_t = \text{iTransformer}(E_s) \in \mathbb{R}^{B \times C \times L}, \label{eq_transformer}
\end{equation}
where $B$ is the batch size. Unlike traditional Transformers that apply attention across time steps, iTransformer applies self-attention to the channel dimension, capturing spatial relationships between brain regions while preserving temporal information. This spatial modeling approach is particularly beneficial for subsequent frequency-domain analysis, as the enhanced inter-channel relationships learned by iTransformer remain invariant during FFT transformation. Since FFT operates independently on each channel's temporal sequence, the spatial connectivity patterns captured by iTransformer are preserved across different frequency bands, ensuring that both spatial and spectral information are effectively integrated without functional conflict.

\subsubsection{Hierarchical Mutual-Cross Attention Block}

The core innovation of ABEMA lies in its hierarchical mutual-cross attention mechanism, which integrates complementary features from different frequency bands. Emotional states are closely associated with specific EEG frequency band activities; for instance, alpha bands correlate with relaxation states, while beta bands relate to alertness and concentration. Our approach captures complex relationships both within and between frequency bands through a two-level attention structure.

\textbf{Frequency Band Decomposition and Feature Extraction:} We first transform the temporal EEG signals into the frequency domain using Fast Fourier Transform (FFT):
\begin{equation}
	\varGamma = \mathcal{F}[E_{\text{t}}] \in \mathbb{R}^{B \times C \times F}, \label{eq_frequency}
\end{equation}
where $B$ is the batch size, $C$ is the number of channels, and $F$ is the number of frequency points. We then extract signals from five standard frequency bands using band selection masks:
\begin{align}
	\varGamma_{b} &= \varGamma \odot M_b(f), \quad b \in \{\delta, \theta, \alpha, \beta, \gamma\}
\end{align}
where $M_b(f)$ is the band selection mask, and $\delta$(0.5-4Hz), $\theta$(4-8Hz), $\alpha$(8-13Hz), $\beta$(13-30Hz), and $\gamma$(30-50Hz) represent the five standard EEG frequency bands. The frequency-domain decomposition in ABEMA provides inherent robustness against residual EMG artifacts, as speech-related muscle activities primarily contaminate frequencies above 30Hz \cite{muthukumaraswamy2013high} while emotion-relevant EEG signals concentrate in lower frequency bands. Compared to raw temporal signals, band-specific masking structurally confines such high-frequency contamination to the $\gamma$ band, leaving the $\delta$, $\theta$, $\alpha$, and $\beta$ bands largely unaffected.

DE and PSD features computed as band-level statistical summaries inherently suppress the transient spike characteristics of EMG bursts, yielding more stable representations under active speech conditions. For each band, we extract DE and PSD:
\begin{align}
    D_b &= \frac{1}{2}\log(2\pi e\sigma_b^2) \in \mathbb{R}^{B \times C} \\
    P_b &= \frac{1}{F}\sum_{f \in b} |\varGamma_f|^2 \in \mathbb{R}^{B \times C}
\end{align}
where $\sigma_b^2$ is the variance of the signal in band $b$. DE measures signal complexity and uncertainty, while PSD quantifies energy distribution across frequencies, jointly providing comprehensive emotional state representation.

\textbf{Intra-band Mutual-Cross Attention:} For each frequency band, we adopt the mutual-cross attention mechanism \cite{zhao2024feature} to fuse DE and PSD features. We extend this into a hierarchical structure, first applying it within frequency bands and subsequently establishing connections between bands. We generate query (Q), key (K), and value (V) matrices according to the following equations:
\begin{equation}
\begin{aligned}
    Q_b^D &= D_b \mathbf{W}_b^{QD}, \quad K_b^P = P_b \mathbf{W}_b^{KP}, \quad V_b^P = P_b \mathbf{W}_b^{VP}, \\
    Q_b^P &= P_b \mathbf{W}_b^{QP}, \quad K_b^D = D_b \mathbf{W}_b^{KD}, \quad V_b^D = D_b \mathbf{W}_b^{VD},
\end{aligned}
\end{equation}
where $\mathbf{W}_b^{QD}$, $\mathbf{W}_b^{KP}$, $\mathbf{W}_b^{VP}$, $\mathbf{W}_b^{QP}$, $\mathbf{W}_b^{KD}$, $\mathbf{W}_b^{VD} \in \mathbb{R}^{C \times d_k}$ are learnable parameter matrices, and $d_k$ is the hidden dimension. Then, we calculate bidirectional mutual-cross attention:
\begin{equation}
    A_b^{DP} = \text{softmax}\left(\frac{Q_b^D K_b^{P^T}}{\sqrt{d_k}}\right)V_b^P,  A_b^{PD} = \text{softmax}\left(\frac{Q_b^P K_b^{D^T}}{\sqrt{d_k}}\right)V_b^D,
\end{equation}

Finally, we combine the outputs from both directions:
\begin{equation}
    F_b = A_b^{DP} + A_b^{PD} \in \mathbb{R}^{B \times d_k},
\end{equation}

\textbf{Inter-band Mutual-Cross Attention:} 
After intra-band feature fusion, we design a second layer of mutual-cross attention to model relationships between frequency bands.  Emotional states typically manifest as coordinated changes across multiple frequency bands with significant individual variations. Our inter-band attention mechanism adapts to these individual differences through independent parameter optimization for each subject, enabling the discovery of subject-specific optimal frequency interaction patterns rather than enforcing identical frequency relationships across all subjects.

Firstly, we concatenate the fused features from all frequency bands as follow:
\begin{equation}
    F_{all} = [F_{\delta}; F_{\theta}; F_{\alpha}; F_{\beta}; F_{\gamma}] \in \mathbb{R}^{B \times 5d_k}
\end{equation}
Then, we generate query vectors for each frequency band and shared key and value matrices:
\begin{align}
    Q_b^{f} &= F_b \mathbf{W}_b^{Qf} \in \mathbb{R}^{B \times d_k} \\
    K^{f} &= F_{all} \mathbf{W}^{Kf}, \quad V^{f} = F_{all} \mathbf{W}^{Vf} \in \mathbb{R}^{B \times d_k}
\end{align}
where $\mathbf{W}_b^{Qf} \in \mathbb{R}^{d_k \times d_k}$ and $\mathbf{W}^{Kf}, \mathbf{W}^{Vf} \in \mathbb{R}^{5d_k \times d_k}$ are learnable parameter matrices. These parameters are optimized independently for each subject, allowing the attention mechanism to learn personalized frequency interaction patterns that reflect individual neurophysiological characteristics.

We calculate the attention of each band to all bands:
\begin{equation}
    A_b^{f} = \text{softmax}\left(\frac{Q_b^{f} (K^{f})^T}{\sqrt{d_k}}\right)V^{f} \in \mathbb{R}^{B \times d_k}
\end{equation}

Finally, we integrate the original features and attention outputs through residual connections:
\begin{equation}
    \hat{F}_b = F_b + A_b^{f} \in \mathbb{R}^{B \times d_k}
\end{equation}

The residual connection preserves original frequency-specific information while incorporating cross-band interactions. This design allows each individual's model to balance between preserving their unique frequency characteristics and learning optimal cross-frequency relationships, with the learned attention patterns reflecting both universal neurophysiological principles and individual neural response variations.

\textbf{Adaptive Band Fusion and Feature Reconstruction:} 
We calculate the importance of each frequency band through a learnable weight vector:
\begin{equation}
    \alpha_b = \frac{\exp(\mathbf{w}^T \hat{F}_b)}{\sum_{i \in \{\delta, \theta, \alpha, \beta, \gamma\}} \exp(\mathbf{w}^T \hat{F}_i)},
\end{equation}
where $w \in \mathbb{R}^{d_k}$ is a learnable weight vector. This weight vector is optimized for each subject individually, enabling personalized frequency band prioritization that adapts to individual neurophysiological response patterns. We introduce a balance parameter $\alpha \in [0,1]$ to control the proportion between processed features and the original signal:
\begin{equation}
    E_n = \alpha \cdot \text{LN}\left(\mathcal{F}^{-1}\left(\sum_b \alpha_b \hat{F}_b\right)\right) + (1-\alpha) \cdot E_t,
\end{equation}
where $\mathcal{F}^{-1}$ represents the inverse FFT and LN denotes layer normalization. The resulting $E_{n} \in \mathbb{R}^{B \times C \times L}$ serves as an optimized EEG embedding for subsequent multi-modal fusion. The adaptive fusion mechanism creates subject-specific frequency representations by dynamically weighting different frequency contributions, addressing individual variations in frequency-emotion mapping patterns while maintaining the same architectural framework across all subjects.

\subsubsection{Feature Normalization and Projection}
After adaptive band fusion, we apply layer normalization to the EEG features and linear projection to map the normalized features to a shared embedding space, generating EEG embeddings that align dimensionally with audio and video features for subsequent hypergraph multi-modal fusion.

\subsection{Unimodal Encoders}

To enhance the representation capability of individual modality features and facilitate subsequent cross-modal information fusion, we encode each modality into a unified d-dimensional semantic space using specialized strategies tailored for different modality characteristics.

Following common practice in multi-modal learning, we employ fully connected networks for audio and video features, while for EEG modality, we adopt Bidirectional Gated Recurrent Units (BiGRU) to capture bidirectional temporal dependencies due to the complexity of temporal characteristics and dynamic nature of emotional state changes. The specific encoding process is formulated as:
\begin{equation}
    \begin{aligned}
    v_i^a &= \mathbf{W}_a A_{ij} + b_a \in \mathbb{R}^d, \\
    v_i^v &= \mathbf{W}_v V_{ij} + b_v \in \mathbb{R}^d, \\
    v_i^e &= \text{BiGRU}(E_n, v_{i(+,-)}^e) \in \mathbb{R}^d
\end{aligned}
\end{equation}
where $W_a \in \mathbb{R}^{d \times d_a}$ and $W_v \in \mathbb{R}^{d \times d_v}$ are the learnable weight matrices for audio and video modalities respectively, with corresponding bias vectors $b_a$ and $b_v$. The raw audio features $A_{ij}$ are extracted using the openSMILE toolkit with the IS10 configuration \cite{eyben2010opensmile} from the audios, while the raw  facial expressions features $V_{ij}$ are extracted using a pre-trained MA-NET \cite{zhao2021learning} from the videos. For EEG encoding, $E_n$ represents the EEG features processed by ABEMA, and $v_{i(+,-)}^e$ represents the bidirectional processing result that fuses forward and backward temporal information.

This differentiated encoding strategy ensures that features from each modality are mapped into a unified semantic space while preserving their inherent characteristics, providing a solid foundation for subsequent multi-modal fusion.

\subsection{Adaptive Hypergraph Fusion Module (AHFM)}

Current approaches to ERC often simplify cross-modal interactions by modeling them as pairwise relationships (e.g., audio-text or video-text). In our study, our AHFM uses hypergraphs to directly capture complex higher-order relationships (e.g., simultaneous EEG-audio-video dependencies), which better reflect the group dynamics of multi-modal emotional cues. Furthermore, since each modality contributes uniquely to detecting instantaneous emotional shifts, we integrate learnable modality-specific weights. These weights are dynamically adjusted during training to prioritize the most informative modalities.

\subsubsection{Hypergraph Construction}

We employ hypergraph theory to model complex relational structures in multi-modal conversations. Given a conversation sequence with N utterance segments, we construct a hypergraph $\mathcal{G} = (\mathcal{V}, \mathcal{E})$ to represent multi-modal interaction patterns. The node set $\mathcal{V} = \{v_i^x | i \in [1,N], x \in \{e,a,v\}\}$ contains all unimodal segments, where $v_i^e$, $v_i^a$, $v_i^v$ correspond to EEG, audio, and video modalities of the $i$-th segment respectively, totaling $3N$ nodes.

Based on the principles of multi-modal synergy and temporal continuity in emotional expression, we design two types of hyperedge connection patterns to capture different levels of relational dependencies. Intra-segment hyperedges $e_i^s = \{v_i^e, v_i^a, v_i^v\}$ connect all modality nodes within the same segment, modeling the synergistic effects of instantaneous multi-modal emotional expression and forming the intra-segment hyperedge set $\mathcal{E}_s$ with $|\mathcal{E}_s| = N$. Inter-segment hyperedges $e_x^t = \{v_i^x | i \in [1,N]\}$ connect nodes of the same modality across different time segments, capturing temporal evolution patterns of emotional states and constituting the inter-segment hyperedge set $\mathcal{E}_t$ with $|\mathcal{E}_t| = 3$. This design enables the hypergraph to simultaneously model synergistic relationships between modalities and evolutionary relationships across time, totaling $N+3$ hyperedges.

To effectively model the differentiated contributions of nodes in different hyperedge types, we introduce an adaptive weight allocation mechanism with a node weight function $\omega: \mathcal{V} \times \mathcal{E} \rightarrow \mathbb{R}^+$ to quantify the influence intensity of node $v_i^x$ within a specific hyperedge $e_j$.

We design a hierarchical weighting strategy where intra-segment hyperedges $e_i^s \in \mathcal{E}_s$ use node weights $\alpha_s^x$ to reflect the synergistic contributions of different modalities in instantaneous emotional expression, while inter-segment hyperedges $e_x^t \in \mathcal{E}_t$ use weights $\alpha_t^x$ to capture the continuity strength of the same modality in temporal evolution. This constructs the weighted incidence matrix $\mathbf{\hat{H}} \in \mathbb{R}^{|\mathcal{V}| \times |\mathcal{E}|}$:

\begin{equation}
    \mathbf{\hat{H}}_{ij} = 
    \begin{cases} 
    \alpha_s^x, & \text{if } v_i^x \in e_j^s \text{ and } e_j^s \in \mathcal{E}_s; \\ 
    \alpha_t^x, & \text{if } v_i^x \in e_j^t \text{ and } e_j^t \in \mathcal{E}_t; \\ 
    0, & \text{otherwise.}
    \end{cases}
\end{equation}

Additionally, we define the hyperedge weight diagonal matrix $\mathbf{W}_e = \text{diag}(\beta_s^1, ..., \beta_s^N, \beta_t^1, \beta_t^2, \beta_t^3)$, where $\beta_s^i$ and $\beta_t^j$ represent the importance weights of intra-segment and inter-segment hyperedges respectively, enabling adaptive learning of optimal weight distributions for each modality under different relational patterns.

\subsubsection{Multi-modal Hypergraph Fusion}

Based on the successful application of hypergraph convolution \cite{chen2023multivariate}, we adapt it to multi-modal EEG emotion recognition tasks and optimize the information propagation mechanism by integrating our proposed hierarchical weighting strategy. This fusion process gradually refines higher-order multi-modal and contextual relationships through iterative node-hyperedge information exchange. Specifically, we implement information propagation through $L$-layer iterative updates:

\begin{equation}
    \begin{aligned}
        V^{(1)} &= \sigma\left(\mathbf{D}_{\mathcal{V}}^{-1}\mathbf{I} \mathbf{W}_e \mathbf{D}_{\mathcal{E}}^{-1} \mathbf{\hat{H}}^T V^{(0)} \right) \\
    &\dots \\
    V^{(L)} &= \sigma\left(\mathbf{D}_{\mathcal{V}}^{-1}\mathbf{I} \mathbf{W}_e \mathbf{D}_{\mathcal{E}}^{-1} \mathbf{\hat{H}}^T V^{(L-1)} \right)
    \end{aligned}
\end{equation}
where $V^{(l)} = \{v_{i,(l)}^x|i \in [1,N], x \in \{e,a,v\}\} \in \mathbb{R}^{|\mathcal{V}| \times d}$ is the input feature at layer $l$, $\sigma$ is the non-linear activation function, $\mathbf{D}_{\mathcal{V}} \in \mathbb{R}^{|\mathcal{V}| \times |\mathcal{V}|}$ and $\mathbf{D}_{\mathcal{E}} \in \mathbb{R}^{|\mathcal{E}| \times |\mathcal{E}|}$ are the node degree matrix and hyperedge degree matrix respectively, used for feature normalization. Finally, we concatenate the representations of three modalities to obtain the fused feature:
\begin{equation}
    f_i = [v_{i,(L)}^e; v_{i,(L)}^a; v_{i,(L)}^v]
\end{equation}

\subsection{Emotion Classification}

To achieve final emotion prediction, we employ a multi-layer perceptron classifier to process the fused multi-modal features $f_i$. This classifier adopts a two-layer fully connected structure that can effectively learn complex emotional patterns. Specifically, the classification process includes non-linear transformation in the hidden layer, probability computation in the output layer, and final prediction decision:

\begin{equation}
    \begin{aligned}
h_i &= \text{ReLU}(\mathbf{W}_{c} f_i + \mathbf{b}_{c}) \\
P_i &= \text{softmax}(\mathbf{W}_{o} h_i + \mathbf{b}_{o}) \\
\hat{y}_i &= \arg\max_{c} P_i[c]
\end{aligned}
\end{equation}
where $\mathbf{W}_{c}$ and $\mathbf{b}_{c}$ are the weight matrix and bias vector of the classification layer respectively, $\mathbf{W}_{o}$ and $\mathbf{b}_{o}$ are the parameters of the output layer, and $P_i \in \mathbb{R}^C$ represents the probability distribution over $C$ emotion categories.

During training stage, we employ categorical cross-entropy loss combined with L2 regularization:
\begin{equation}
    \mathcal{L} = -\frac{1}{\sum_{s=1}^{N} |D_s|} \sum_{s=1}^{N} \sum_{i=1}^{|D_s|} \log P_{s,i}[y_{s,i}] + \lambda \|\theta\|_2
\end{equation}
where $N$ represents the number of dialogues, $|D_s|$ denotes the number of utterance segments in the $s$-th dialogue, $P_{s,i}$ is the probability distribution, $y_{s,i}$ is the ground truth label, $\lambda$ is the regularization weight, and $\theta$ represents the trainable parameters of the model.

\section{Experiments}\label{Experiments}

\subsection{EEG-based ERC Dataset}

\begin{table}[t]
    \centering
    \caption{The basic situation and the division of training set and test set of EAV and AFFEC datasets. }\label{tab:Datasets}
    \small
    \setlength{\tabcolsep}{3pt}
    \begin{tabularx}{\textwidth}{>
    {\centering\arraybackslash}X>
    {\centering\arraybackslash}X>
    {\centering\arraybackslash}X>
    {\centering\arraybackslash}X>
    {\centering\arraybackslash}X>
    {\centering\arraybackslash}X
    }
    \toprule
    Datasets & Subjects & Time Split & Partition & Count  \\
    \hline
    \multirow{2}{*}{EAV} & \multirow{2}{*}{42} & \multirow{2}{*}{5s} & train & 11760  \\
                         &                     &                     & test & 5040  \\
    \hline
    \multirow{2}{*}{AFFEC} & \multirow{2}{*}{72} & \multirow{2}{*}{3s} & train & 4480  \\
                         &                     &                     & test & 1152 \\

    \toprule
    
    \end{tabularx}
\end{table}

\textbf{EAV}\cite{lee2024eav}: The recently released multi-modal dialogue emotion dataset, EAV, includes EEG data from 30 channels, audio recordings, and facial expression videos from 42 subjects. 
This dataset represents the first publicly available collection that integrates EEG, audio, and video in a conversational context. Each subject engaged in 200 interactions within prompt-based dialogue scenarios, eliciting five distinct emotions: Neutral, Anger, Happy, Sad and Calm. Each interaction consisted of 20 seconds of listening followed by 20 seconds of speaking. 
For our evaluation, we focused exclusively on the speaking data of the subjects and followed the authors' preprocessing methods, segmenting the 20-second speech data stream into 5-second intervals. 
This approach aims to simulate the interruptions in conversation flow that individuals might encounter due to health conditions or hardware/software issues. By doing so, it disrupts the complete semantic structure and reflects scenarios in which the text modality may be missing or incomplete.

\textbf{AFFEC}\cite{sekiavandi2025advancing}: The Advancing Face-to-Face Emotion Communication (AFFEC) dataset is a multi-modal dataset designed to capture the dynamic complexities of face-to-face emotional interactions by integrating various modalities, including EEG, eye-tracking, GSR, facial actions, and Big Five personality assessment. 
The AFFEC dataset includes a total of 72 participants, covering educational levels from high school to doctoral degrees to ensure participant diversity. The dataset contains 84 simulated dialogues targeting six different emotions (anger, disgust, fear, happiness, neutrality, and sadness), totaling over 5,000 trials with continuous emotion annotation, which is subsequently discretized into three-level categories (High, Medium, Low) for both arousal and valence dimensions in the actual emotion recognition tasks. For our evaluation, we utilized three modalities: EEG, eye-tracking, and GSR. 
The AFFEC dataset originally contains 73 participants, but only 72 participants have available EEG data for download. Therefore, our study utilized 72 participants, covering educational levels from high school to doctoral degrees to ensure participant diversity. The dataset contains 84 simulated dialogues targeting six different emotions (anger, disgust, fear, happiness, neutrality, and sadness), totaling over 5,000 trials with continuous emotion annotation, which is subsequently discretized into three-level categories (High, Medium, Low) for both arousal and valence dimensions in the actual emotion recognition tasks. For our evaluation, we utilized three modalities: EEG, eye-tracking, and GSR.
To address the challenge of limited sample size per subject for subject-wise analysis, we implemented a time windowing strategy that segments each trial into overlapping time windows, effectively expanding the dataset by a factor of five while preserving the temporal dynamics of emotional states. Importantly, to avoid data leakage, we performed the training-testing split at the subject level before applying the time windowing strategy. This data augmentation approach ensures sufficient training samples for robust subject-specific model learning while maintaining the integrity of physiological signal patterns.

\begin{table}[t]
    \centering
    \caption{Hyperparameter configurations for EAV and AFFEC datasets.}
    \label{tab:hyperparameters}
    \small
    \setlength{\tabcolsep}{4pt}
    \begin{tabularx}{\textwidth}{>{\centering\arraybackslash}X>{\centering\arraybackslash}X>{\centering\arraybackslash}X>{\centering\arraybackslash}X>{\centering\arraybackslash}X>{\centering\arraybackslash}X>{\centering\arraybackslash}X>{\centering\arraybackslash}X}
    \toprule
     & Batch & lr & $D_h$ & L & Dropout  & Epochs & $\alpha$ \\
    \hline
    EAV & 32 & 1e-4 & 512 & 3 & 0.5  & 30 & 0.7\\
    AFFEC & 16 & 1e-4 & 512 & 3 & 0.6  & 30 & 0.7 \\
    \toprule
    \end{tabularx}
\end{table}

\begin{table}[t]
    \centering
    \caption{Subject-wise performance of Hyper-MML on EAV dataset. }\label{tab:Subject-Wise-result}
    \small
    \setlength{\tabcolsep}{3pt}
    \begin{tabularx}{\textwidth}{>
    {\centering\arraybackslash}X>{\centering\arraybackslash}X>{\centering\arraybackslash}X>{\centering\arraybackslash}X>
    {\centering\arraybackslash}X>
    {\centering\arraybackslash}X>
    {\centering\arraybackslash}X>
    {\centering\arraybackslash}X>
    {\centering\arraybackslash}X
    }
    \toprule
    Subject & Acc& F1& Subject& Acc& F1 & Subject& Acc& F1\\
    \hline

    1 & 67.50 & 67.15 & 15 & 86.67 & 86.63 & 29 & 69.17 & 68.15 \\
    2 & 91.67 & 91.71 & 16 & 70.83 & 70.17 & 30 & 80.83 & 81.30 \\
    3 & 85.83 & 86.06 & 17 & 99.17 & 99.17 & 31 & 85.00 & 84.60 \\
    4 & 83.33 & 82.17 & 18 & 76.67 & 76.44 & 32 & 65.00 & 63.96\\
    5 & 70.83 & 69.62 & 19 & 71.67 & 71.94 & 33 & 83.33 & 82.99\\
    6 & 80.00 & 79.58 & 20 & 92.50 & 92.46 & 34 & 70.83 & 69.79\\
    7 & 81.67 & 81.44 & 21 & 86.67 & 86.05 & 35 & 67.50 & 67.55 \\
    8 & 74.17 & 74.80 & 22 & 82.50 & 82.51 & 36 & 74.17 & 73.97\\
    9 & 83.33 & 83.18 & 23 & 77.50 & 76.66 & 37 & 62.50 & 61.96\\
    10 & 70.00 & 66.16 & 24 & 85.83 & 85.66 & 38 & 85.83 & 85.98\\
    11 & 75.00 & 74.46 & 25 & 74.17 & 74.68 & 39 & 76.67 & 75.24\\
    12 & 76.67 & 75.81 & 26 & 70.83 & 70.17 & 40 & 70.83 & 70.89\\
    13 & 84.17 & 84.12 & 27 & 85.83 & 85.27 & 41 & 76.67 & 76.27 \\
    14 & 70.00 & 68.71 & 28 & 83.33 & 83.99 & 42 & 78.33 & 77.98\\
    \toprule
    & & & & & & Average & 78.21& 77.80 \\
    \cline{7-9}
    
    \end{tabularx}
\end{table}

\begin{table*}[t]
    \small
    \caption{Performance comparison between our method and other competing methods on EAV dataset across five emotion categories (Neutral, Anger, Happy, Sad, Calm). Statistical significance: * p < 0.05  compared to the best baseline. The best result in each column is presented in \textbf{bold}.}\label{tab:EAVresult}
    \centering
    \setlength{\tabcolsep}{3pt}
    \begin{tabularx}{\textwidth}{>
    {\centering\arraybackslash}p{3cm}>
    {\centering\arraybackslash}p{1.3cm}>
    {\centering\arraybackslash}p{1.3cm}>
    {\centering\arraybackslash}p{1.3cm}>
    {\centering\arraybackslash}p{1.3cm}>{\centering\arraybackslash}p{1.3cm}>
    {\centering\arraybackslash}X
    }
        \toprule
        \multirow{3}{*}{Methods} & \multicolumn{6}{c}{EAV} \\
        \cline{2-7}
        & \multicolumn{5}{c}{Emotion Categories (F1)} & Overall \\
        \cline{2-7}
        & Neutral & Anger & Happy & Sad & Calm & Acc~~~~F1\\
        \hline
         bc-LSTM\cite{ma2019emotion} & 58.16 & 64.57 & 47.13 & 64.48 & 52.14 & 57.21 ~57.30\\
         DialogueRN\cite{majumder2019dialoguernn} & 63.75 & 67.12 & 44.16 & 65.19 & 65.50 & 61.20 ~61.14\\
         DialogueGCN\cite{ghosal2019dialoguegcn} & 68.32 & 70.18 & 56.17 & 80.14 & 68.94 & 68.73 ~68.75\\
         MMGCN\cite{hu2021mmgcn} & 72.88 & \textbf{74.12} & 55.38 & 81.28 & 73.01 & 71.31 ~71.33\\
         MM-DFN\cite{hu2022mm}\ & 71.89 & 70.34 & 59.15 & 74.42 & 71.92 & 69.74 ~69.54\\
         GraphMFT\cite{li2023graphmft} & 71.71 & 72.91 & 60.15 & 82.66 & 72.51 & 71.55 ~72.99\\
         M3NET\cite{mane2020multi} & 73.45 & 71.94 & 79.36 & 81.64 & 70.72 & 75.14 ~75.42\\
         HAUCL\cite{yi2024multimodal} & 75.21 & 72.80 & 82.44 & 78.17 & 70.97 & 75.91 ~75.92\\
         AGF-IB\cite{shou2024adversarial} & 73.53 & 71.76 & 81.87 & 80.91 & 72.34 & 76.19 ~76.08\\
         
         \hline

         MMResLSTM\cite{ma2019emotion} & 70.25 & 68.42 & 65.18 & 78.35 & 71.75 & 72.37 ~72.19\\
         HetEmotionNet\cite{jia2021hetemotionnet} & 72.18 & 70.85 & 69.42 & 80.12 & 73.68 & 74.70 ~74.25\\
         MDNet\cite{jia2024multi} & 74.92 & 73.28 & 75.15 & 82.47 & \textbf{75.08} & 76.92 ~76.78\\

        AMERL\cite{yin2025eeg} & 75.18 & 71.88 & 72.80 & 79.25 & 72.62 & 74.82 ~74.35\\

         \hline
         Hyper-MML & \textbf{76.39} & 73.05 & \textbf{82.89}* & \textbf{84.55}* & 72.14 & \textbf{78.21}* ~\textbf{77.80*}\\
         \toprule
    \end{tabularx}
\end{table*}

\begin{table*}[t]
    \small
    \centering
    \caption{Performance comparison between our method and other competing methods on AFFEC dataset for arousal and valence classification tasks. Statistical significance: * p < 0.05  compared to the best baseline. The best result in each column is presented in in \textbf{bold}.}\label{tab:AFFECresult}
    \begin{tabularx}{\textwidth}{>
    {\centering\arraybackslash}p{3.5cm}>
    {\centering\arraybackslash}X>
    {\centering\arraybackslash}X>
    {\centering\arraybackslash}X>
    {\centering\arraybackslash}X
    }
         \toprule
         \multirow{2}{*}{Methods} & \multicolumn{4}{c}{AFFEC task} \\
         \cline{2-5}
         & Perceived-Arousal & Perceived-Valence & Felt-Arousal& Felt-Valence \\
         \hline
         bc-LSTM\cite{ma2019emotion} & 35.21 & 31.45 & 42.67 & 38.92 \\
         DialogueRNN\cite{majumder2019dialoguernn} & 38.94 & 33.78 & 45.83 & 41.26 \\
         DialogueGCN\cite{ghosal2019dialoguegcn} & 43.25 & 37.89 & 50.74 & 46.18 \\
         MMGCN\cite{hu2021mmgcn} & 46.83 & 40.56 & 53.27 & 49.71 \\
         MM-DFN\cite{hu2022mm} & 45.92 & 39.84 & 52.45 & 48.89 \\
         GraphMFT\cite{li2023graphmft} & 47.68 & 41.87 & 54.69 & 51.02 \\
         M3NET\cite{mane2020multi} & 49.34 & 43.21 & 56.85 & 52.78 \\
         HAUCL\cite{yi2024multimodal} & 50.87 & 44.67 & 58.42 & 54.31 \\
         AGF-IB\cite{shou2024adversarial} & 51.23 & 45.12 & 58.91 & 54.87 \\
         \hline
         MMResLSTM\cite{ma2019emotion} & 42.85 & 36.72 & 49.41 & 45.18 \\
         HetEmotionNet\cite{jia2021hetemotionnet} & 46.29& 39.84 & 53.16 & 48.92 \\
         MDNet\cite{jia2024multi} & 50.12 & 43.67 & 57.38 & 53.25 \\
         AMERL\cite{yin2025eeg} & 50.45 & 44.38 & 57.96 & 53.94 \\
         
         \hline
         
         Hyper-MML & \textbf{53.76}* & \textbf{47.89}* & \textbf{60.23}* & \textbf{57.32}* \\
         \toprule
    \end{tabularx}
\end{table*}

\begin{table}[t]
    \centering
    \caption{Subject-wise performance of Hyper-MML on AFFEC dataset. }\label{tab:Subject-Wise-result—affec}
    \small
    \setlength{\tabcolsep}{3pt}
    \begin{tabularx}{\textwidth}{>
    {\centering\arraybackslash}X>{\centering\arraybackslash}X>{\centering\arraybackslash}X>{\centering\arraybackslash}X>
    {\centering\arraybackslash}X>
    {\centering\arraybackslash}X>
    {\centering\arraybackslash}X>
    {\centering\arraybackslash}X>
    {\centering\arraybackslash}X
    }
    \toprule
    Subject & Acc & F1& Subject& Acc& F1 & Subject& Acc& F1\\
    \hline
    1 & 62.12 & 58.34 & 25 & 45.45 & 41.23 & 49 & 68.94 & 65.78 \\
    2 & 73.48 & 71.92 & 26 & 59.09 & 55.67 & 50 & 47.73 & 43.89 \\
    3 & 56.06 & 52.41 & 27 & 71.97 & 69.85 & 51 & 65.15 & 62.03 \\
    4 & 68.18 & 65.29 & 28 & 53.03 & 49.17 & 52 & 60.61 & 57.24 \\
    5 & 44.70 & 40.83 & 29 & 66.67 & 63.51 & 53 & 72.73 & 70.18 \\
    6 & 58.33 & 54.76 & 30 & 49.24 & 45.67 & 54 & 55.30 & 51.42 \\
    7 & 63.64 & 60.29 & 31 & 70.45 & 67.83 & 55 & 67.42 & 64.15 \\
    8 & 51.52 & 47.89 & 32 & 46.97 & 43.21 & 56 & 61.36 & 58.07 \\
    9 & 69.70 & 66.94 & 33 & 64.39 & 61.12 & 57 & 73.48 & 71.26 \\
    10 & 57.58 & 53.85 & 34 & 52.27 & 48.54 & 58 & 48.48 & 44.73 \\
    11 & 65.91 & 62.67 & 35 & 71.21 & 68.45 & 59 & 62.88 & 59.41 \\
    12 & 50.76 & 46.92 & 36 & 45.45 & 41.67 & 60 & 56.82 & 53.18 \\
    13 & 67.42 & 64.28 & 37 & 59.85 & 56.23 & 61 & 69.70 & 66.87 \\
    14 & 54.55 & 50.84 & 38 & 73.48 & 71.15 & 62 & 51.52 & 47.64 \\
    15 & 71.21 & 68.73 & 39 & 47.73 & 43.95 & 63 & 64.39 & 61.28 \\
    16 & 46.21 & 42.18 & 40 & 61.36 & 58.19 & 64 & 58.33 & 54.89 \\
    17 & 60.61 & 57.12 & 41 & 55.30 & 51.73 & 65 & 72.73 & 70.04 \\
    18 & 68.94 & 65.87 & 42 & 68.18 & 65.42 & 66 & 46.97 & 43.35 \\
    19 & 53.79 & 50.16 & 43 & 72.73 & 70.11 & 67 & 60.61 & 57.48 \\
    20 & 65.15 & 61.89 & 44 & 50.00 & 46.25 & 68 & 54.55 & 51.07 \\
    21 & 49.24 & 45.38 & 45 & 63.64 & 60.47 & 69 & 67.42 & 64.52 \\
    22 & 71.97 & 69.12 & 46 & 57.58 & 53.96 & 70 & 52.27 & 48.81 \\
    23 & 45.45 & 41.74 & 47 & 70.45 & 67.59 & 71 & 65.91 & 62.84 \\
    24 & 59.09 & 55.43 & 48 & 48.48 & 44.61 & 72 & 59.85 & 56.37 \\
    \toprule
     & & & & & & Average & 60.23 & 56.78 \\
    \cline{7-9}
    \end{tabularx}
\end{table}

\subsection{Implementation Details and Evaluation Metrics}

For both EAV and AFFEC datasets, we adopted a subject-wise splitting strategy to ensure complete independence between training and testing sets. The data splitting was performed at the subject level before any temporal segmentation or data augmentation procedures. The basic situation and division details of the EAV and AFFEC datasets are shown in Table \ref{tab:Datasets}.

Our experiments were conducted on a Windows 11 system equipped with an NVIDIA RTX 3090 GPU. The framework was implemented using Python 3.8 and PyTorch 1.7.1. We employed the Adam optimizer for parameter updates with a learning rate of 0.0001. 
Training was conducted for 30 epochs with early stopping based on validation performance to prevent overfitting. The model parameters were initialized using Xavier initialization for linear layers and orthogonal initialization for recurrent components. The detailed hyperparameter configurations for both datasets are presented in Table \ref{tab:hyperparameters}. 

For ABEMA module, the frequency bands were set to standard EEG ranges: $\delta$ (0.5-4Hz), $\theta$ (4-8Hz), $\alpha$ (8-13Hz), $\beta$ (13-30Hz), and $\gamma$ (30-50Hz). The attention mechanism employed 6 attention heads. The balance parameter $\alpha$ in adaptive band fusion was set to 0.7. For AHFM module, the hidden dimension $D_h$ determines the feature representation capacity across all modalities, while the number of hypergraph layers $L$ controls the depth of higher-order relationship modeling. Node weights and hyperedge weights were initialized uniformly in the range [0.1, 1.0] and optimized during training.

Model performance was evaluated using two standard metrics: accuracy (Acc) and F1-score (F1). Accuracy measures the proportion of correctly classified samples, while F1-score provides a balanced assessment considering both precision and recall, particularly important for imbalanced emotion datasets.

\subsection{Competing Methods}

In order to validate the effectiveness of our proposed Hyper-MML for multi-modal ERC task, we conducted extensive comparative experiments. However, there is currently insufficient exploration in the field of dialogue emotion recognition based on EEG. Aside from one multi-modal dialogue emotion recognition model utilizing the EEG modality, known as AMERL, there are virtually no other direct benchmark models available in the literature. Therefore, we selected several emotion recognition benchmark models that are widely applied in the field of Natural Language Processing (NLP), which typically integrate text, audio, and facial expression modalities for multi-modal emotion analysis. Additionally, we incorporated three traditional EEG-based multi-modal physiological signal fusion models for emotion recognition to better evaluate the potential of EEG signals. For these benchmark models, we replaced the text modality embeddings with EEG modality embeddings to assess the potential of EEG signals in emotion recognition.  Below, we present a brief introduction to some of the comparative models employed in our study:

\textbf{bc-LSTM}\cite{ma2019emotion}: A bidirectional LSTM variant that processes utterances in both temporal directions, capturing contextual dependencies for improved sentiment classification in video sequences.

\textbf{DialogueRNN}\cite{majumder2019dialoguernn}: This RNN-based architecture models participant states and contextual dynamics using gated recurrent units, effectively tracking emotional evolution between speakers and listeners.

\textbf{DialogueGCN}\cite{ghosal2019dialoguegcn}: The first graph convolutional approach for conversational emotion recognition, addressing context propagation limitations in RNN methods by modeling self-dependency and inter-speaker relationships.

\textbf{MMGCN}\cite{hu2021mmgcn}: A deep graph convolutional framework that fuses multi-modal information while capturing long-range contextual dependencies and inter-speaker interactions in dialogue sequences.

\textbf{MM-DFN}\cite{hu2022mm}: This dynamic fusion network integrates text, audio, and video through interactive multiview memory modules, adapting feature importance based on conversational context.

\textbf{GraphMFT}\cite{li2023graphmft}: A graph attention-based technique that simultaneously captures intra-modal context and inter-modal complementarity through enhanced attention mechanisms.

\textbf{M3NET}\cite{mane2020multi}: This approach models multivariate relationships and multi-frequency signals using graph neural networks, capturing complex utterance interdependencies in conversational scenarios.

\textbf{HAUCL}\cite{yi2024multimodal}: A hypergraph-based framework combining variational autoencoders with contrastive learning to dynamically adjust connections while reducing contextual redundancy and over-smoothing.

\textbf{AGF-IB}\cite{shou2024adversarial}: This method eliminates inter-modal heterogeneity through information bottleneck theory and adversarial learning, while employing graph contrastive learning for semantic information capture.

\textbf{MMResLSTM}\cite{ma2019emotion}: A multi-modal residual LSTM network that learns temporal correlations among EEG and physiological signals through weight sharing across modalities with spatial-temporal shortcut paths.

\textbf{HetEmotionNet}\cite{jia2021hetemotionnet}: A two-stream heterogeneous graph recurrent neural network that models heterogeneity and correlation among multi-modal physiological signals through graph transformer and convolutional networks.

\textbf{MDNet}\cite{jia2024multi}: A multi-level disentangling network that models consistency and heterogeneity of multi-modal physiological signals through modality-level and subject-level disentangling modules for cross-subject emotion recognition.

\textbf{AMERL}\cite{yin2025eeg}: A multi-modal framework specifically designed for multi-modal integration with EEG signals, using dynamic attention mechanisms to adaptively weight EEG, video, image, and audio features for robust emotion recognition.

\section{Results and Discussion}\label{Results and Discussion}

\subsection{Comparison with Baselines}

Table \ref{tab:Subject-Wise-result} presents the subject-wise performance of our proposed Hyper-MML framework across all 42 participants in the EAV dataset. The results demonstrate considerable inter-subject variability, with accuracy ranging from 62.50\% to 99.17\%, reflecting the inherent individual differences in EEG signal characteristics. Despite this variability, our method achieves an overall average accuracy of 78.21\% and F1-score of 77.80\%, indicating robust performance in EEG-based conversational emotion recognition.

Table \ref{tab:EAVresult} provides a comprehensive comparison between our Hyper-MML framework and other baseline methods on the EAV dataset. The baseline methods span from traditional RNN-based approaches (bc-LSTM, DialogueRNN) to advanced graph-based methods (DialogueGCN, MMGCN, GraphMFT), recent hypergraph techniques (M3NET, HAUCL, AGF-IB) and traditional EEG-based multi-modal methods (MMResLSTM, HetEmotionNet, MDNet). Our approach demonstrates superior performance across multiple evaluation metrics, achieving the highest overall accuracy of 78.21\% and F1-score of 77.80\%, representing significant improvements of  1.29\% in accuracy and 1.02\% in F1-score compared to the previous best-performing method MDNet. Statistical significance testing confirms that these improvements are statistically meaningful (p < 0.05 for overall performance metrics). The performance gains are particularly pronounced in the recognition of specific emotional categories: our method achieves the highest F1-scores for Sad (84.55\%) and Happy (82.89\%) emotions, with improvements of  2.08\% and 7.74\% respectively over the previous best results.

While hypergraph-based methods generally outperform traditional approaches, our method achieves superior results through more effective utilization of the hypergraph structure rather than increased model complexity. Our superior performance stems from two key factors rather than architectural complexity. Multi-modal integration with EEG signals provides direct neural correlates, offering more reliable indicators than reconstructed features in competing methods. Additionally, AHFM explicitly models both intra-segment multi-modal synergy and inter-segment temporal continuity, while existing methods focus primarily on complex fusion without optimizing hypergraph structure.

To validate the generalizability of our approach, we further evaluated Hyper-MML on the AFFEC dataset, which presents a more challenging task of arousal and valence classification in three-level categories (Low/Medium/High). Table \ref{tab:AFFECresult} demonstrates that our method consistently outperforms all baseline approaches across four distinct classification tasks, achieving the highest F1-scores of 53.76\% for Perceived-Arousal, 47.89\% for Perceived-Valence, 60.23\% for Felt-Arousal, and 57.32\% for Felt-Valence, representing improvements of  3.31\%, 2.77\%, 1.32\%, and 2.45\% respectively over the previous best-performing methods. All improvements achieve statistical significance (p < 0.05), confirming the robustness of our approach across different emotional dimensions. Results confirm established patterns: felt emotions outperform perceived emotions, and arousal tasks exceed valence tasks in accuracy. Table \ref{tab:Subject-Wise-result—affec} presents the subject-wise performance across all 72 participants for Felt-Valence task on the AFFEC dataset, showing an overall average accuracy of 60.23\% and F1-score of 56.78\% with considerable inter-subject variability.

\subsection{Effectiveness of ABEMA}

\begin{table}[t]
    \centering
    \caption{\small Comparison of EEG encoding methods on EAV and AFFEC datasets. The results of AFFEC come from Felt-Arousal task. The best result in each column is presented in \textbf{bold}.}\label{tab:encoder}
    
    \begin{tabularx}{\textwidth}{>
    {\centering\arraybackslash}p{3cm}>
    {\centering\arraybackslash}X>
    {\centering\arraybackslash}X>
    {\centering\arraybackslash}X>
    {\centering\arraybackslash}X
    }
        \toprule
        \multirow{2}{*}{Encoder} & \multicolumn{2}{c}{EAV} & \multicolumn{2}{c}{AFFEC}  \\
        \cline{2-5}
         & Acc & F1 & Acc & F1  \\
        \hline
        MLP & 61.80 & 61.56 & 55.97 & 55.67   \\
        EEGNet\cite{lawhern2018eegnet} & 69.18 & 69.94 & 58.08 & 54.84   \\
        TSConv\cite{song2023decoding} & 73.22 & 72.12 & 58.12 & 54.67   \\
        ATMS\cite{li2024visual} & 76.12 & 75.83 & 59.12 & 55.08   \\
        NESTA\cite{li2024neural} & 76.40 & 76.37 & 59.80 & 55.82   \\
        \hline
        ABEMA & \textbf{78.21} & \textbf{77.80} & \textbf{60.23} & \textbf{56.78}  \\
        \toprule
    \end{tabularx}
    
\end{table}

As shown in Table \ref{tab:encoder}, our proposed ABEMA encoder demonstrates superior performance compared to five established EEG processing methods across both datasets. On the EAV dataset, ABEMA achieves the highest accuracy of 78.21\% and F1-score of 77.80\%, outperforming the second-best method NESTA by 1.81\% in accuracy and 1.43\% in F1-score respectively. The comparative results clearly illustrate the evolution from traditional MLP approaches with 61.80\% accuracy to advanced deep learning methods like EEGNetV4 and TSConv achieving 69.18\% and 73.22\% accuracy respectively, with attention-based methods ATMS and NESTA showing further improvements. ABEMA's hierarchical mutual-cross attention mechanism and adaptive frequency band fusion strategy achieve the best results by effectively modeling both intra-band and inter-band relationships in EEG signals.

We also report results from the Felt-Arousal classification task on the AFFEC dataset. AFFEC results maintain this advantage with 60.23\% accuracy, exceeding NESTA by 0.43\%. This consistent advantage across different task formulations demonstrates that ABEMA's subject-specific adaptation layer and hierarchical attention mechanism effectively address fundamental challenges in EEG-based emotion recognition, including inter-subject variability and complex frequency-domain relationships.

Additionally, compared to raw temporal signal, ABEMA's frequency-domain processing provides architectural-level robustness against high-frequency EMG artifacts induced by facial muscle activity during active speech, as emotion-relevant neural oscillations are structurally isolated in lower frequency bands.

\subsection{Modality Contribution Analysis}

\begin{table}[t]
    \centering
    \small
    \caption{Modality contribution analysis on EAV dataset comparing individual and combined modality performance. The best result in each column is presented in \textbf{bold}.}\label{tab:modalities}
    \begin{tabularx}{\textwidth}{>
    {\centering\arraybackslash}p{5cm}>
    {\centering\arraybackslash}X>
    {\centering\arraybackslash}X
    }
        \toprule
        \multirow{2}{*}{Modalities} & \multicolumn{2}{c}{EAV}  \\
        \cline{2-3}
         & Acc & F1 \\
        \hline
        EEG & 71.83 & 72.04  \\
        Audio(OpenSMile) & 69.47 & 69.88   \\
        Audio(AST) & 69.26 & 69.31 \\ 
        Video(MANET) & 68.36 & 68.54   \\
        Video(Vivit) & 68.14 & 69.21   \\
        \hline
        EEG+Audio+Video & \textbf{78.21} & \textbf{77.80} \\
        \toprule
    \end{tabularx}
\end{table}

\begin{figure}
    \centering
    \includegraphics[width=0.5\linewidth]{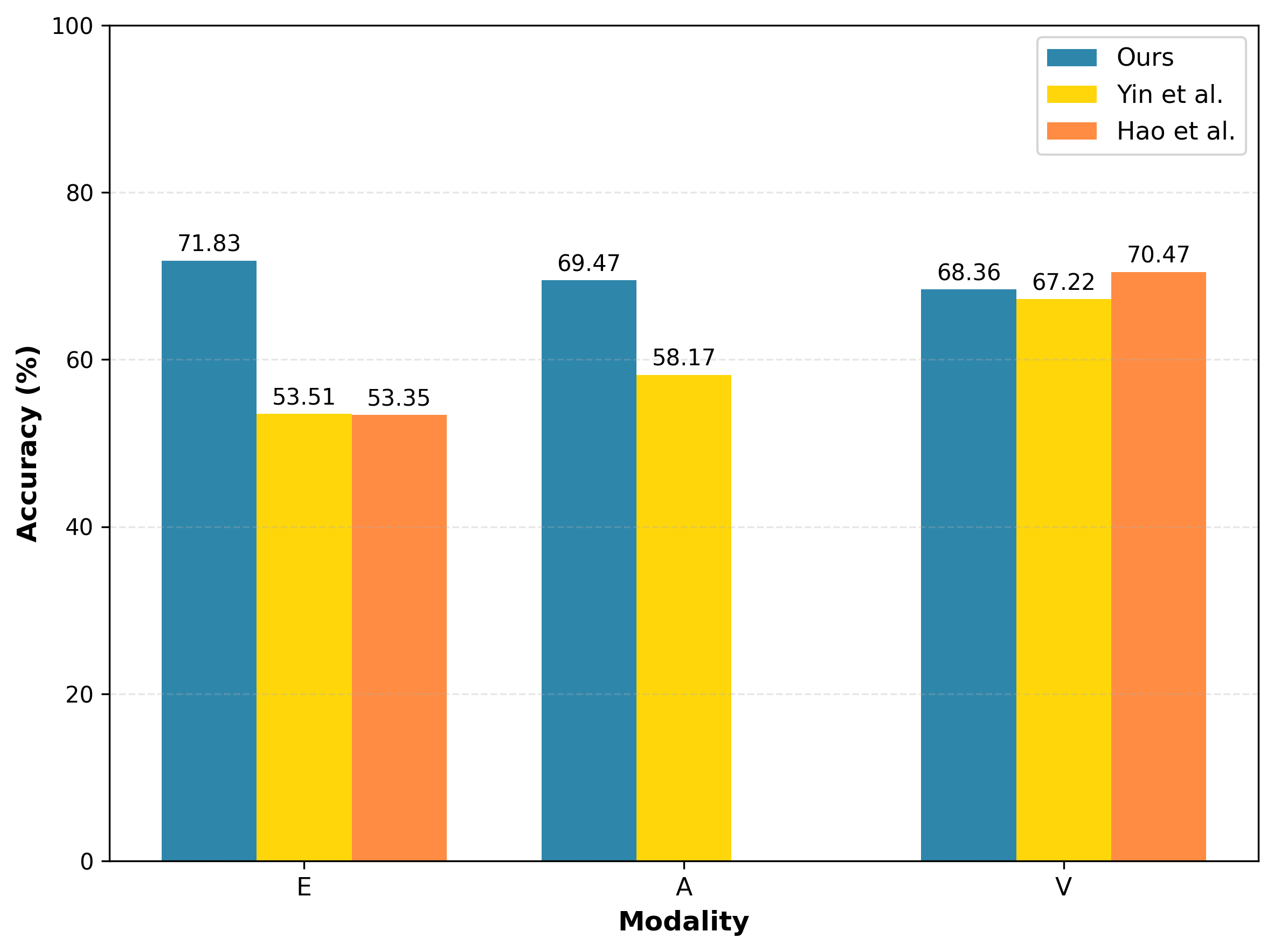}
    \caption{Comparison of single modality performance across different studies on the EAV dataset. E, A, and V represent EEG, Audio, and Video modalities respectively.}
    \label{fig:modality_comparison}
\end{figure}

Table \ref{tab:modalities} presents a comprehensive analysis of individual modality contributions in our multi-modal framework. It is important to note that these single-modality results are derived from embeddings that have undergone information exchange through our hypergraph fusion framework, reflecting each modality's representational capacity after inter-modal interactions. Among the three modalities, EEG demonstrates the strongest individual performance with 71.83\% accuracy and 72.04\% F1-score, highlighting the critical role of physiological signals in capturing objective emotional states. For audio modality, we compare two feature extraction approaches: OpenSMILE features achieve 69.47\% accuracy and 69.88\% F1-score, while the Transformer-based AST\cite{gong21b_interspeech} achieves 69.26\% accuracy and 69.31\% F1-score, demonstrating comparable performance across different feature extraction methods. Similarly, for video modality, MA-NET features achieve 68.36\% accuracy and 68.54\% F1-score, while ViViT\cite{arnab2021vivit} achieves 68.14\% accuracy and 69.21\% F1-score, showing consistent performance levels that reflect the inherent challenges of extracting subtle emotional cues from facial expressions in conversational contexts.

The integration of all three modalities (EEG+Audio+Video) achieves substantial performance improvements, reaching 78.21\% accuracy and 77.80\% F1-score, representing gains of 6.38\% in accuracy and 5.76\% in F1-score compared to the best individual modality (EEG). This significant enhancement demonstrates the complementary nature of different modalities in capturing the multifaceted aspects of emotional expression. As shown in Figure \ref{fig:modality_comparison}, the observed hierarchical performance of modalities (i.e., EEG > Audio $\approx$ Video) in our conversational setting contrasts with recent studies \cite{yin2025eeg, hao2025step} where video modality typically demonstrates superior performance. This discrepancy can be attributed to some key factors in our experimental setup and methodology: (1) Data characteristics and experimental paradigm differences: Hao et al.\cite{hao2025step} likely utilized the passive listening portion of the EAV dataset containing only EEG-video modalities, where facial expressions remain uncontaminated by speech-related muscular activities. However, our study focuses on active speaking scenarios where facial muscle movements during natural conversation interfere with facial expression recognition, thereby possibly reducing video modality accuracy \cite{Mariooryad2015facial} while EEG signals capture genuine emotional neural responses; (2) EEG processing procedure: while Hao et al.\cite{hao2025step} employed simple time window-based EEG encoders and Yin et al.\cite{yin2025eeg} used standard transformer-based encoders, our ABEMA module incorporates subject-specific calibration, frequency-domain artifact mitigation, and hierarchical attention mechanisms specifically optimized for conversational EEG processing, enabling superior extraction of emotion-relevant neural patterns compared to generic approaches.

These results further confirm that in active conversational environments, facial muscle movements significantly impact video modality recognition accuracy, while EEG demonstrates superior advantages in such scenarios due to its direct capture of neural emotional responses. While EEG maintains strong representational capacity even after hypergraph updates due to its rich neural information, audio and video modalities benefit more significantly from multi-modal collaboration, as evidenced by their relatively lower individual performance but substantial contribution to the combined result. The superior performance of the combined approach validates our hypergraph-based fusion strategy's effectiveness in modeling complex higher-order relationships among modalities, enabling the framework to leverage the unique strengths of each modality while compensating for their individual limitations.

\subsection{Hypergraph Structure Superiority Analysis}

\begin{table*}[t]
    \centering
    \caption{Graph structure comparison on EAV and AFFEC datasets. Statistical significance: * p < 0.05 compared to Multi-GCN. The best result in each column is presented in \textbf{bold}.}
    \label{tab:graph_structure_comparison}
    \small
    \begin{tabularx}{\textwidth}{>{\centering\arraybackslash}p{3cm}>{\centering\arraybackslash}X>{\centering\arraybackslash}X>{\centering\arraybackslash}X>{\centering\arraybackslash}X}
        \toprule
        \multirow{3}{*}{Methods} & \multicolumn{2}{c}{EAV} & \multicolumn{2}{c}{AFFEC (Felt-Arousal)} \\
        \cline{2-5}
        & Acc (\%) & F1 (\%) & Acc (\%) & F1 (\%) \\
        \hline
        GCN-Pairwise & 74.85 & 74.12 & 56.89 & 53.24 \\
        GAT-Pairwise & 75.92 & 75.48 & 57.65 & 54.12 \\
        Multi-GCN & 76.34 & 76.01 & 58.41 & 54.89 \\
        \hline
        Hyper-MML & \textbf{78.21}* & \textbf{77.80}* & \textbf{60.23}* & \textbf{56.78}* \\
        \toprule
    \end{tabularx}
\end{table*}

To validate the superiority of hypergraph structures over traditional graph-based approaches, we conducted comprehensive comparisons with three representative graph neural network variants: GCN-Pairwise models second-order relationships through pairwise connections (e.g., EEG-Audio, Audio-Video), GAT-Pairwise enhances pairwise modeling with attention mechanisms but remains limited to second-order interactions, and Multi-GCN attempts to approximate higher-order relationships through multiple layers of pairwise connections. In contrast, our Hyper-MML directly models higher-order relationships using hyperedges that simultaneously connect EEG, audio, and video modalities. As shown in Table \ref{tab:graph_structure_comparison}, Hyper-MML consistently outperforms all traditional approaches, achieving 78.21\% accuracy on EAV dataset with 1.87\% improvement over Multi-GCN, and 60.23\% accuracy on AFFEC dataset with 1.82\% improvement.

The superior performance demonstrates that direct hypergraph modeling of triadic EEG-audio-video relationships captures the synchronized multi-modal patterns inherent in emotional expression more effectively than sequential pairwise approximations. While traditional graph methods decompose complex emotional interactions into chains of binary relationships, losing critical information about simultaneous multi-modal synergies, our hypergraph structure preserves the natural higher-order dependencies that characterize conversational emotion recognition. This architectural advantage enables more accurate modeling of the complex dynamics where physiological responses, vocal patterns, and facial expressions exhibit coordinated activation patterns during emotional states.

\subsection{Ablation Study}

\begin{table}[t]
    \centering
    \small
    \caption{Ablation study results on EAV dataset examining key components of ABEMA and AHFM modules. The best result in each column is presented in \textbf{bold}.}\label{tab:Ablation}
    \begin{tabularx}{\textwidth}{>
    {\centering\arraybackslash}p{3cm}>
    {\centering\arraybackslash}p{6cm}>
    {\centering\arraybackslash}X>
    {\centering\arraybackslash}X
    }
        \toprule
        \multicolumn{2}{c}{Component Settings} & Acc & F1 \\
        \hline
        \multirow{4}{*}{ABEMA} & w/o Subject Layer & 73.52 & 73.98   \\
        & w/o Intra-band Attention & 72.11 & 72.28   \\
         & w/o Inter-band Attention & 72.29 & 72.83   \\
         & w/o Both Attention & 70.24 & 70.67   \\
        \hline
        \multirow{3}{*}{AHFM} & w/o Node Weights ($\alpha_s^x$, $\alpha_t^x$)& 75.92 & 75.49   \\
         & w/o Hyperedge Weights ($\beta_s^i$, $\beta_t^j$) & 75.22 & 75.31  \\
         & w/o Both Weights & 72.46 & 72.87  \\
        \hline
        \multicolumn{2}{c}{\textbf{Hyper-MM}L} & \textbf{78.2}1 & \textbf{77.80}  \\
        \hline
    \end{tabularx}
\end{table}

To validate the effectiveness of key components in our Hyper-MML framework, we perform comprehensive ablation experiments on the key components of Hyper-MML in Table \ref{tab:Ablation}.

\textbf{Performance Evaluation of ABEMA:} The ablation study on ABEMA reveals the critical importance of  both the subject-specific layer and the hierarchical mutual-cross attention mechanism. Removing the subject-specific layer results in 73.52\% accuracy and 73.98\% F1-score, representing performance drops of 4.69\% and 3.82\% respectively, confirming its essential role in handling inter-subject variability. Removing intra-band attention results in 72.11\% accuracy and 72.28\% F1-score, while removing inter-band attention leads to 72.29\% accuracy and 72.83\% F1-score, indicating that inter-band relationships are slightly more crucial than intra-band dependencies. The most significant performance degradation occurs when both attention mechanisms are removed, dropping to 70.24\% accuracy and 70.67\% F1-score, representing decreases of 7.97\% and 7.13\% respectively. This substantial decline confirms that the hierarchical attention design effectively captures complex frequency-domain relationships in EEG signals, enabling optimal integration of complementary information across different frequency bands.

\textbf{Performance Evaluation of AHFM:} The AHFM ablation experiments demonstrate the effectiveness of the adaptive weighting strategy in hypergraph construction. Removing node weights results in 75.92\% accuracy and 75.49\% F1-score, while removing hyperedge weights leads to 75.22\% accuracy and 75.31\% F1-score, suggesting that node-level adaptive weighting has a slightly greater impact on performance. When both weight mechanisms are eliminated, performance drops to 72.46\% accuracy and 72.87\% F1-score, representing decreases of 5.75\% and 4.93\% respectively. These results validate that the adaptive weight allocation mechanism enables the model to dynamically prioritize the most informative modalities and relationship types, optimizing information propagation within the hypergraph structure and significantly enhancing multi-modal fusion effectiveness.

\begin{figure}
    \centering
    \includegraphics[width=1\linewidth]{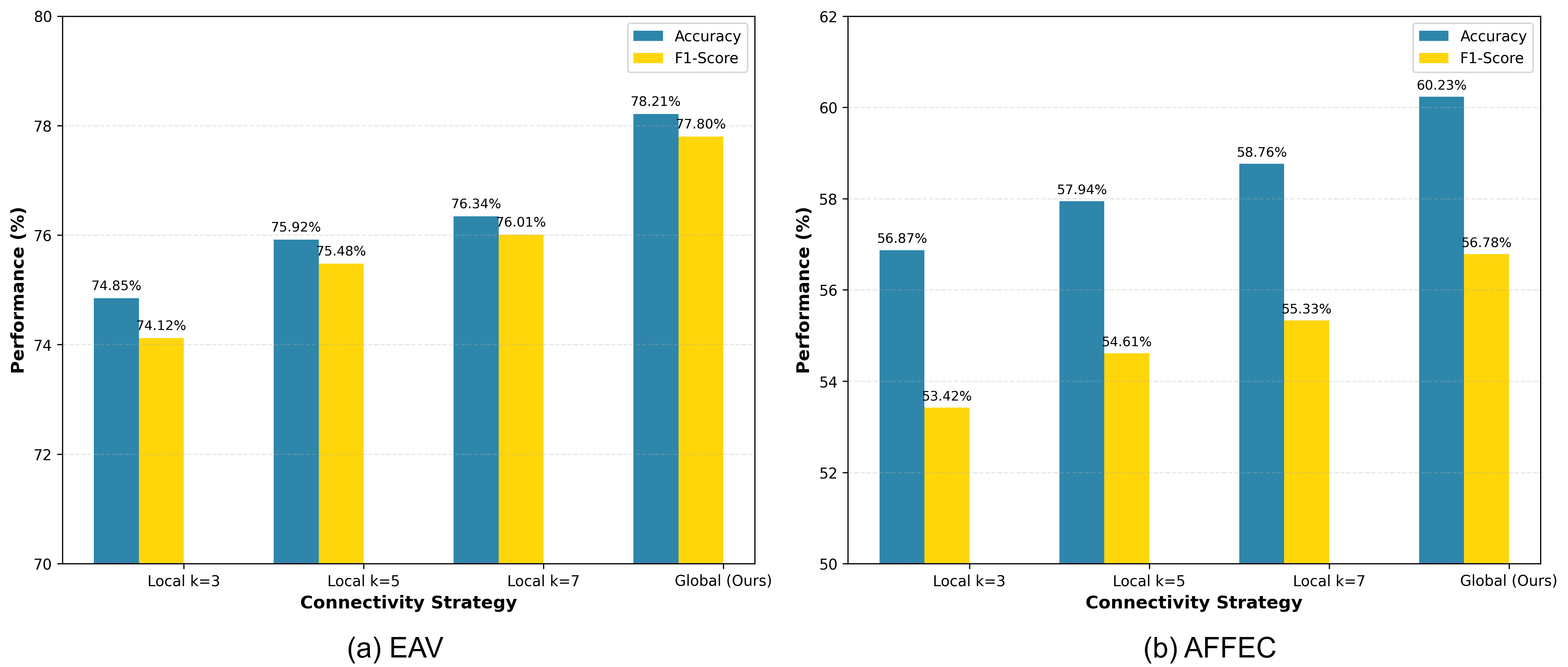}
    \caption{Performance comparison of different hypergraph connectivity strategies on EAV and AFFEC datasets. Global connectivity consistently outperforms local connectivity variants across both datasets, demonstrating the importance of modeling long-range temporal dependencies in conversational emotion recognition.}
    \label{fig:connectivity_comparison}
\end{figure}

\textbf{Hypergraph Connectivity Analysis:} To validate our global inter-segment connectivity strategy, we compare it against local connectivity approaches with different window sizes (k=3, 5, 7) on both datasets. As shown in Figure \ref{fig:connectivity_comparison}, global connectivity consistently outperforms all local variants across both EAV and AFFEC datasets. The performance gap demonstrates that emotional states in conversations exhibit long-range dependencies that extend beyond immediate temporal neighborhoods. Local connectivity with smaller windows fails to capture delayed emotional responses and cross-temporal correlations, while larger windows show gradual improvement but remain suboptimal. Global connectivity enables comprehensive modeling of emotional evolution patterns across entire conversations, validating our design choice for capturing complex temporal dynamics in conversational emotion recognition.

\begin{table}[t]
    \centering
    \small
    \caption{Ablation study results on EAV dataset examining subject-specific layer. The best result in each column is presented in \textbf{bold}.}\label{tab:linearVSnonlinear}
    \begin{tabularx}{\textwidth}{>
    {\centering\arraybackslash}p{4cm}>
    {\centering\arraybackslash}p{5cm}>
    {\centering\arraybackslash}X>
    {\centering\arraybackslash}X
    }
        \toprule
        \multicolumn{2}{c}{Subject Layer Component Settings} & Acc & F1 \\
        \hline
        \multirow{3}{*}{Nonlinear Transformation} & Linear+ReLU & 77.04 & 76.91   \\
        & Linear+GELU & 77.32 & 77.21   \\
         & MLP & 76.75 & 76.83   \\
         \hline
        Linear Transformation & \textbf{Linear(Ours)} & \textbf{78.21} & \textbf{77.80}   \\
        \hline
    \end{tabularx}
\end{table}

\subsection{Subject-specific layer Analysis}

In order to verify the effectiveness of linear transformation for handling individual EEG variations, we conducted a comprehensive ablation study comparing our linear subject-specific layer against various nonlinear alternatives. As shown in Table \ref{tab:linearVSnonlinear}, our linear transformation achieves favorable performance with 78.21\% accuracy and 77.80\% F1-score, compared to Linear+ReLU (77.04\% accuracy), Linear+GELU (77.32\% accuracy), and MLP (76.75\% accuracy). While nonlinear processing has demonstrated advantages in various contexts, these results suggest that linear transformation represents a more suitable approach for addressing inter-subject variability in our Hyper-MML model for the EEG-based conversational emotion recognition task.

The superior performance of linear transformation can be attributed to several interconnected factors. The primary purpose of the subject-specific layer is to perform participant-specific linear calibration rather than complex feature transformation. Research \cite{li2023cross} has demonstrated  that inter-subject variability in EEG signals primarily manifests as linear scaling differences, spatial offset variations, and linear channel mixing effects due to anatomical differences in brain structure, electrode placement sensitivity, and impedance variations, which can be effectively addressed through linear transformation matrices without requiring nonlinear activations. Further, as the subject-specific layer operates at the earliest stage of our processing pipeline, introducing nonlinear activations at this point can inadvertently suppress subtle neural patterns crucial for emotion recognition, while linear transformation preserves the full dynamic range and temporal characteristics needed for downstream frequency-domain processing. Since subsequent layers in our ABEMA module already incorporate sophisticated nonlinear processing through hierarchical mutual-cross attention mechanisms and frequency-domain transformations, there is no necessity to introduce nonlinear complexity prematurely at the calibration stage. Moreover, in conversational emotion recognition tasks where data availability per subject is inherently limited, nonlinear transformations become particularly susceptible to overfitting. The performance variations observed across different activation functions—with ReLU showing greater degradation than GELU—reflect varying degrees of overfitting phenomena, where nonlinear activations tend to memorize subject-specific noise patterns rather than learning generalizable calibration mappings.

\subsection{Subject-Independent Performance Analysis}

\begin{table}[t]
    \centering
    \caption{Performance comparison between subject-wise and subject-independent evaluation on EAV dataset. Performance drop indicates the degradation from subject-wise to subject-independent settings. Statistical significance: * p
< 0.05 compared to the best baseline.}
    \label{tab:subject_independent}
    \small
    \begin{tabularx}{\textwidth}{>{\centering\arraybackslash}p{3cm}>{\centering\arraybackslash}X>{\centering\arraybackslash}X>{\centering\arraybackslash}X>{\centering\arraybackslash}X}
        \toprule
        \multirow{2}{*}{Methods} & \multicolumn{2}{c}{Subject-wise} & \multicolumn{2}{c}{Subject-independent}  \\
        \cline{2-5}
        & Acc & F1 & Acc & F1 \\
        \hline
        bc-LSTM\cite{ma2019emotion} & 57.21 & 57.30 & 45.87 & 45.94 \\
        DialogueRNN\cite{majumder2019dialoguernn} & 61.20 & 61.14 & 49.76 & 49.68 \\
        MMGCN\cite{hu2021mmgcn} & 71.31 & 71.33 & 59.42 & 59.51  \\
        MM-DFN\cite{hu2022mm} & 69.74 & 69.54 & 58.15 & 57.89  \\
        M3NET\cite{mane2020multi} & 75.14 & 75.42 & 64.87 & 65.12  \\
        HAUCL\cite{yi2024multimodal} & 75.91 & 75.92 & 65.73 & 65.81  \\
        \hline
        MMResLSTM\cite{ma2019emotion} & 72.37 & 72.19 & 64.58 & 64.35 \\
        HetEmotionNet\cite{jia2021hetemotionnet} & 74.70 & 74.25 & 67.92 & 67.41  \\
        MDNet\cite{jia2024multi} & 76.92 & 76.78 & 70.15 & 69.94 \\
        AMERL\cite{yin2025eeg} & 74.82 & 74.35 & 68.34 & 67.89 \\
        \hline
        \textbf{Hyper-MML} & \textbf{78.21*} & \textbf{77.80*} & \textbf{73.96*} & \textbf{73.62*} \\
        \toprule
    \end{tabularx}
\end{table}

To validate the generalizability across unseen subjects, we conducted Leave-One-Subject-Out (LOSO) cross-validation on the EAV dataset. In this protocol, we iteratively train on 41 subjects and test on the remaining held-out subject, repeating for all 42 subjects to simulate real-world deployment scenarios.

Our Hyper-MML framework achieves 73.96\% accuracy and 73.62\% F1-score under subject-independent evaluation, representing a modest degradation of 4.25\% in accuracy and 4.18\% in F1-score compared to subject-wise performance (Table \ref{tab:subject_independent}). This relatively small performance gap demonstrates the effectiveness of our subject-specific layer and multi-modal fusion strategy in handling inter-subject variability.

The results reveal distinct patterns across different method categories. Text-centric approaches show substantial performance drops (8-12\%) due to their reliance on semantic coherence that varies significantly across subjects. In contrast, EEG-based methods including our approach maintain more stable performance, with our framework achieving the best subject-independent results while showing the smallest performance degradation among all competing methods. Notably, when examining single-modality EEG performance under subject-independent evaluation, we observe a more pronounced degradation, highlighting the inherent challenge of cross-subject EEG variability. However, the integration of audio and video modalities provides natural robustness against individual physiological differences, as these behavioral cues exhibit more consistent patterns across subjects compared to neural signals. This multi-modal complementarity explains our framework's superior cross-subject stability, where audio-video information compensates for EEG variability while our subject-specific layer effectively normalizes the remaining individual differences.

\begin{figure*}[t]
    \centering
    \includegraphics[width=0.95\linewidth]{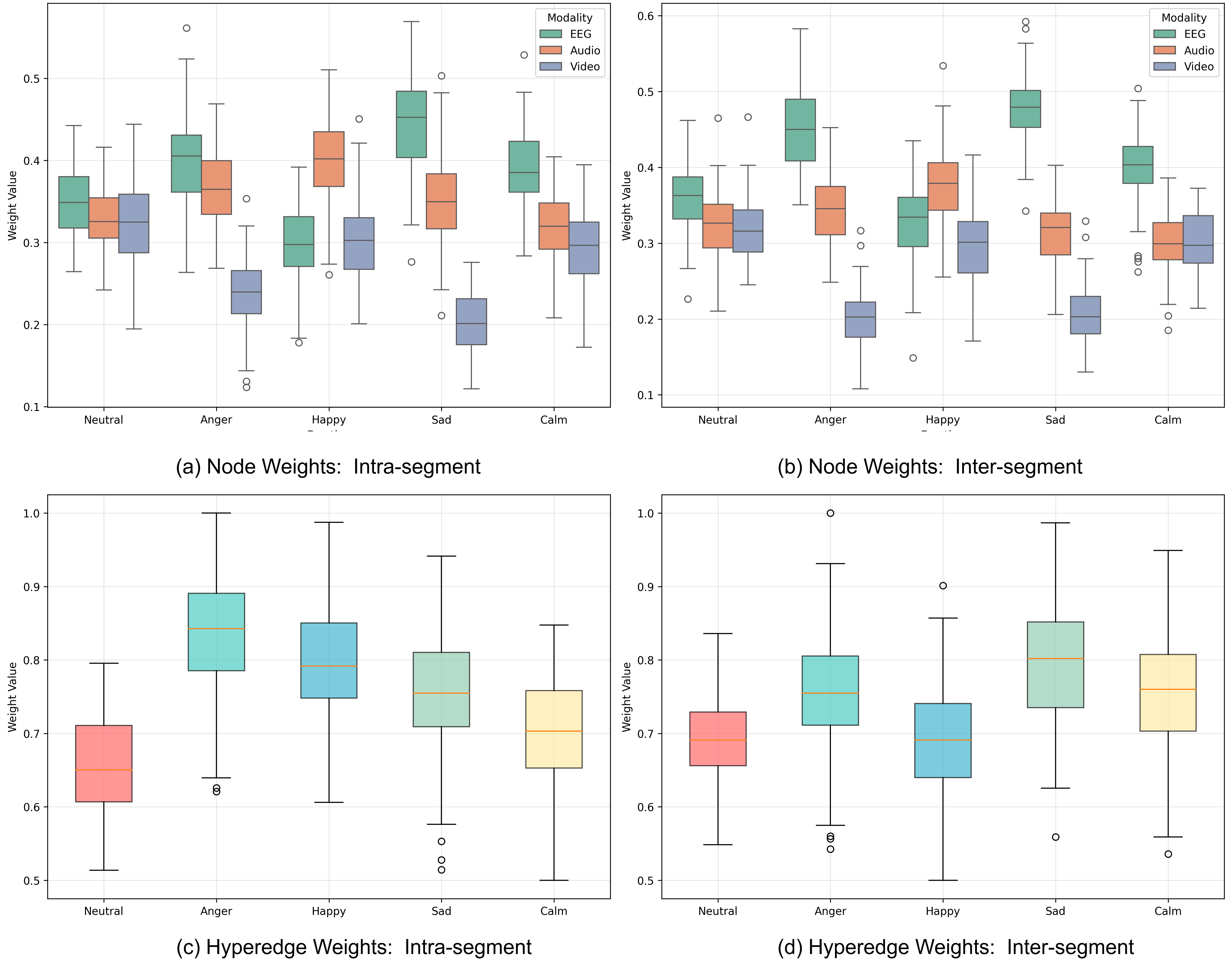}
    \caption{Hypergraph weight distribution analysis across emotions on EAV dataset. (a) Node weights for intra-segment hyperedges ($\alpha_s^x$). (b) Node weights for inter-segment hyperedges ($\alpha_t^x$). (c) Hyperedge weights for intra-segment connections ($\beta_s^i$). (d) Hyperedge weights for inter-segment connections ($\beta_t^j$).}
    \label{fig:weight_analysis}
\end{figure*}

\begin{table}[t]
    \centering
    \caption{Statistical analysis of hypergraph weight distributions across modalities on EAV dataset.}
    \label{tab:weight_statistics}
    
    \small
    \begin{tabularx}{\textwidth}{>{\centering\arraybackslash}p{2cm}>
    {\centering\arraybackslash}p{4cm}>
    {\centering\arraybackslash}X>
    {\centering\arraybackslash}X>
    {\centering\arraybackslash}X>
    {\centering\arraybackslash}X}
    \toprule
    \multirow{2}{*}{Modality} & \multirow{2}{*}{Weight Type} & \multicolumn{4}{c}{Statistical Measures} \\
    \cline{3-6}
    & & Mean & Std & Min & Max \\
    \midrule
    \multirow{2}{*}{EEG} 
    & Intra-segment ($\alpha_s^e$) & 0.376 & 0.070 & 0.178 & 0.569 \\
    & Inter-segment ($\alpha_t^e$) & 0.404 & 0.073 & 0.149 & 0.592 \\
    \midrule
    \multirow{2}{*}{Audio} 
    & Intra-segment ($\alpha_s^a$) & 0.353 & 0.054 & 0.208 & 0.510 \\
    & Inter-segment ($\alpha_t^a$) & 0.332 & 0.051 & 0.185 & 0.534 \\
    \midrule
    \multirow{2}{*}{Video} 
    & Intra-segment ($\alpha_s^v$) & 0.271 & 0.063 & 0.122 & 0.450 \\
    & Inter-segment ($\alpha_t^v$) & 0.265 & 0.064 & 0.108 & 0.466 \\
    \bottomrule
    \end{tabularx}
\end{table}

\begin{table}[t]
    \centering
    \caption{Emotion-specific weight distribution analysis showing modality dominance patterns and hyperedge strength for intra-segment connections on EAV dataset. \textbf{Bold} values indicate the dominant modality for each emotion category.}
    \label{tab:emotion_weight_distribution}
    \small
    \begin{tabularx}
    {\textwidth}{>{\centering\arraybackslash}p{2cm}>
    {\centering\arraybackslash}X>
    {\centering\arraybackslash}X>
    {\centering\arraybackslash}X>
    {\centering\arraybackslash}X}
    
    \toprule
    \multirow{2}{*}{Emotion} & \multicolumn{3}{c}{Modality Weights ($\alpha_s^x$)} & \multirow{2}{*}{\makecell{Hyperedge\\Strength ($\beta_s$)}} \\
    \cline{2-4}
    & EEG & Audio & Video \\
    \midrule
    Neutral & \textbf{0.351} & 0.328 & 0.321 & 0.657 \\
    Anger & \textbf{0.396} & 0.365 & 0.238 & \textbf{0.838} \\
    Happy & 0.299 & \textbf{0.401} & 0.300 & 0.799 \\
    Sad & \textbf{0.446} & 0.352 & 0.202 & 0.752 \\
    Calm & \textbf{0.386} & 0.318 & 0.296 & 0.703 \\
    \bottomrule
    \end{tabularx}
\end{table}

\subsection{Hypergraph Weight Analysis and Interpretability}

To demonstrate the psychological plausibility of our hypergraph structure, we analyzed the learned node weights ($\alpha_s^x$, $\alpha_t^x$) and hyperedge weights ($\beta_s^i$, $\beta_t^j$) across emotion categories in the EAV dataset. Figure \ref{fig:weight_analysis} reveals psychologically meaningful modality prioritization patterns. Tables \ref{tab:weight_statistics} and \ref{tab:emotion_weight_distribution} provide comprehensive statistical analysis of weight distributions, revealing significant patterns that address key questions about modality contributions in conversational emotion recognition.

\textbf{Video's Consistently Low Weight:} Our statistical analysis (Table \ref{tab:weight_statistics}) reveals that video modality demonstrates the lowest mean weights across all conditions (intra-segment:0.271, inter-segment:0.265) with relatively high variability (std=0.063-0.064), indicating unstable contributions. This consistent underperformance possibly stems from three primary factors: (1) Data quality limitations: Natural conversation involves facial muscle movements during speaking that contaminate facial expression recognition, unlike passive viewing scenarios where facial expressions remain uncontaminated\cite{Mariooryad2015facial}; (2) Adaptive fusion mechanism: Our hypergraph structure automatically prioritizes more reliable modalities when video signals exhibit noise, as evidenced by the particularly low video weights in high-intensity emotions like Anger (0.238) and Sad (0.202) where facial muscle tension further degrades expression clarity.

\textbf{EEG's High Weights for Specific Emotions:} Table \ref{tab:emotion_weight_distribution} demonstrates that EEG achieves dominant weights for Neutral(0.351), Anger (0.396), Sadness (0.446), and Calmness (0.386), while showing lower contribution for Happiness (0.299). This pattern reflects distinct neurophysiological characteristics where Anger and Sadness represent high-arousal and internalized emotional states, respectively, both directly captured through beta-wave and altered alpha/theta wave neural oscillations that EEG can objectively measure. Calmness correlates with enhanced alpha-wave dominance indicating parasympathetic activation, making EEG particularly effective for detecting relaxation states that may not manifest clearly through behavioral modalities. In contrast, Happiness exhibits lower EEG weights as this emotion manifests more prominently through external behavioral expressions captured by audio and video modalities. High-arousal emotions show distinct weight distributions: Anger exhibits elevated EEG weights reflecting beta-wave physiological arousal, coupled with substantial Audio weights capturing vocal intensity. Sad demonstrates strong EEG dominance with minimal video contribution, consistent with the internalized nature of depressive states where facial expressions provide limited information. Neutral and Calm emotions display balanced weight distributions, validating adaptive modality weighting based on emotional characteristics.

\textbf{Audio's Superiority for Happiness Recognition:} Table \ref{tab:emotion_weight_distribution} shows that Audio achieves its highest weight specifically for Happiness (0.401), surpassing both EEG (0.299) and Video (0.300). This superiority stems from happiness manifesting through distinctive acoustic properties including clear pitch elevation, rhythm variations, and spontaneous laughter that provide robust and consistent vocal signatures. Unlike facial expressions that can be contaminated during natural conversation, vocal expressions of joy demonstrate greater reliability and resistance to environmental interference while maintaining stable performance across emotions (std=0.051-0.054). Our framework effectively captures these happiness-specific acoustic characteristics such as fundamental frequency increases and spectral brightness changes that distinctively separate positive emotions from neutral or negative states.

Hyperedge weight analysis reveals emotion-specific temporal dynamics. Anger achieves the highest intra-segment weights (0.838), indicating strong instantaneous multi-modal synchronization during emotional peaks. Conversely, Sad exhibits moderate intra-segment weights (0.752), reflecting the sustained but less synchronized nature of depressive states where individual modalities may exhibit temporal delays in emotional expression. These learned weight patterns demonstrate that our adaptive hypergraph structure captures neurophysiologically plausible multi-modal interaction dynamics, with weights automatically adapting to emotion-specific characteristics while maintaining consistency with established emotion psychology principles.

\subsection{Computational Cost and Efficiency Evaluation}

To comprehensively assess the practical applicability of our Hyper-MML framework, we conduct extensive computational cost and efficiency evaluations, comparing against three representative baseline methods across key performance metrics. All measurements are performed on identical hardware configurations (NVIDIA RTX 3090 GPU) using consistent experimental protocols across 10 independent runs.

Table \ref{tab:computational_cost} presents comprehensive computational metrics for all evaluated methods. Despite having the largest parameter count (6.89M), Hyper-MML demonstrates superior efficiency across multiple critical dimensions. Our framework achieves the fastest training time (127s) and convergence speed (12 epochs), representing 18\% and 37\% improvements over the best-performing baselines, respectively, while maintaining competitive inference latency (10.3ms).

\begin{table}[t]
    \centering
    \caption{Computational Cost and Efficiency Evaluation on EAV dataset. Efficiency Score is calculated as $\frac{F1}{Parameters + \frac{Training\_Time}{100} + Latency}$.}
    \label{tab:computational_cost}
    \small
    \setlength{\tabcolsep}{4pt}
    \begin{tabularx}{\textwidth}{>{\centering\arraybackslash}X>{\centering\arraybackslash}X>{\centering\arraybackslash}X>{\centering\arraybackslash}X>{\centering\arraybackslash}X>{\centering\arraybackslash}X>{\centering\arraybackslash}X>{\centering\arraybackslash}X}
    \toprule
    Method & Param- eters(M) & Model Size(MB) & Training Time(s) & Conver-gence(epochs) & Latency(ms) & F1 & Efficiency Score \\
    \hline
    HAUCL\cite{yi2024multimodal} & 5.80 & 23.22 & 165 & 19 & 22.0 & 75.92 & 0.42 \\
    MDNet\cite{jia2024multi} & 4.56 & 18.06 & 155 & 30 & 15.0 & 76.78 & 0.46 \\
    AMERL\cite{yin2025eeg} & \textbf{4.23} & \textbf{16.84} & 200 & 22 & 17.5 & 74.35 & 0.37 \\
    \hline
    HyperMML & 6.89 & 26.30 & \textbf{127} & \textbf{12} & \textbf{10.3} & \textbf{77.80} & \textbf{0.58} \\
    \toprule
    \end{tabularx}
\end{table}

Hyper-MML achieves the highest efficiency score (0.58), demonstrating optimal balance between computational cost and recognition performance. The rapid convergence stems from ABEMA's hierarchical mutual-cross attention enabling efficient frequency-domain feature learning and AHFM's adaptive weight allocation facilitating optimal information propagation across modalities. The inference latency of 10.3ms and model size of 26.30MB remain highly suitable for real-time clinical applications, where diagnostic accuracy takes precedence over computational efficiency.

Future work will explore model compression techniques and knowledge distillation to further reduce computational overhead while maintaining the superior performance characteristics of our hypergraph-based multi-modal fusion approach.

\subsection{Parameter Sensitivity Analysis}

\begin{figure*}[t]
    \centering
    \includegraphics[width=0.95\linewidth]{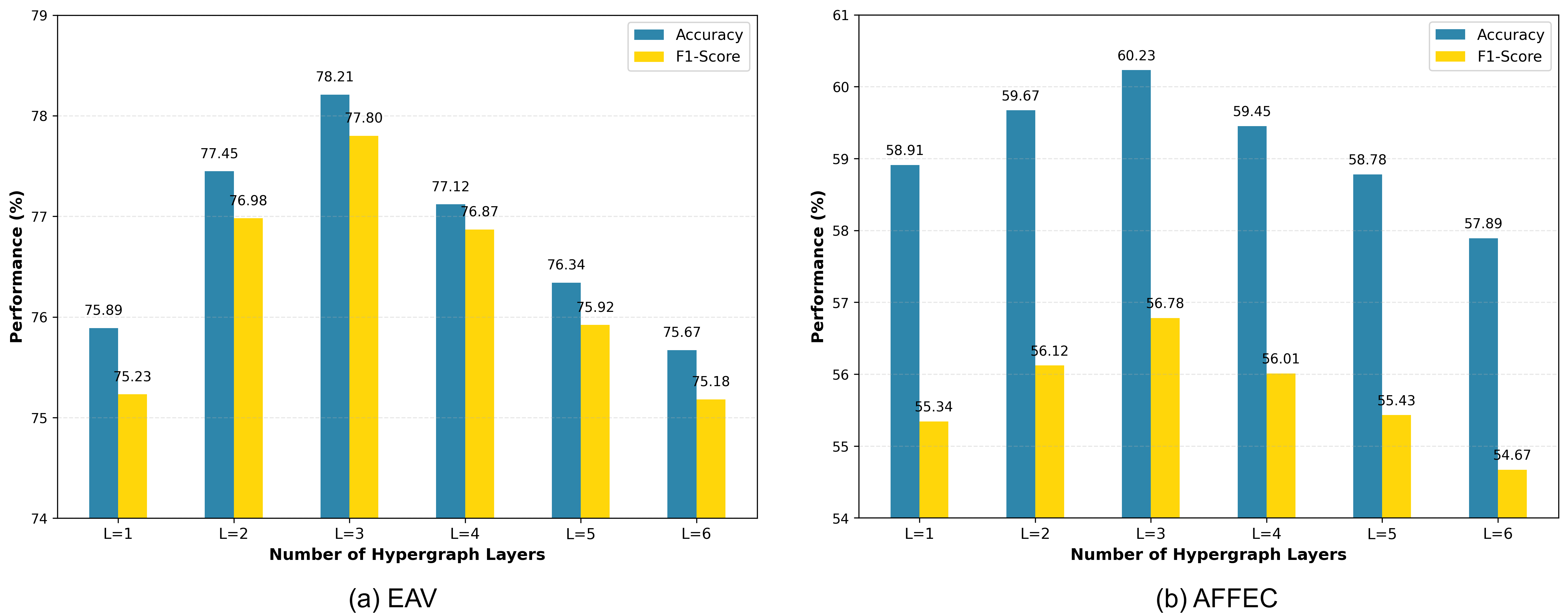}
    \caption{Parameter sensitivity analysis of hypergraph layers (L) on EAV and AFFEC datasets. (a) EAV dataset shows optimal performance at L=3 with gradual degradation for deeper architectures. (b) AFFEC dataset (Felt-Arousal task) demonstrates similar patterns with peak performance at L=3, validating the robustness of our architectural design across different datasets.}
    \label{fig:parameter_sensitivity}
\end{figure*}

To evaluate the robustness and optimal configuration of our Hyper-MML framework, we conducted parameter sensitivity analysis focusing on the number of hypergraph layers (L), a critical architectural component that controls the depth of higher-order relationship modeling. As illustrated in Figure \ref{fig:parameter_sensitivity}, both datasets exhibit consistent patterns with optimal performance achieved at L=3. The results demonstrate that shallow architectures (L=1,2) underutilize the hypergraph's capacity for modeling complex multi-modal interactions, while deeper architectures (L>3) suffer from over-smoothing effects where excessive message passing dilutes distinctive emotional features across nodes. This inverted-U pattern validates our architectural choice and confirms that three hypergraph layers provide the optimal balance between capturing sophisticated multi-modal dependencies and preserving discriminative emotional representations. The consistent trends across both EAV and AFFEC datasets further demonstrate the generalizability of our framework's design principles across different conversational emotion recognition scenarios.

\subsection{Visualization}

\begin{figure}
    \centering
    \includegraphics[width=1\linewidth]{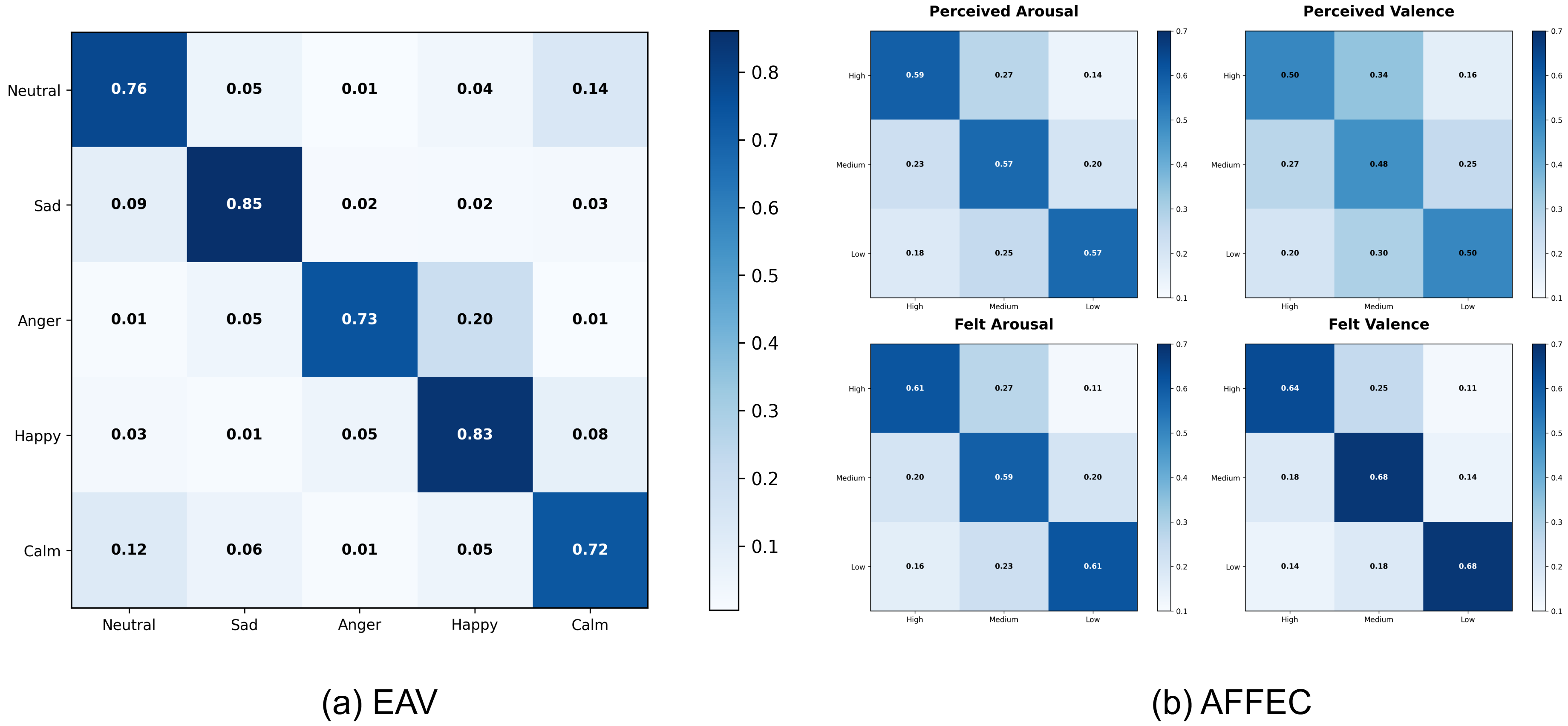}
    \caption{Confusion matrix on the EAV and AFFEC datasets. (a) EAV dataset shows confusion matrix for five emotion categories. (b) AFFEC dataset presents confusion matrices for four classification tasks: Perceived-Arousal, Perceived-Valence, Felt-Arousal, and Felt-Valence with three-level categories (High/Medium/Low).}
    \label{fig:confusion_matrix}
\end{figure}

\begin{figure}
    \centering
    \includegraphics[width=0.95\linewidth]{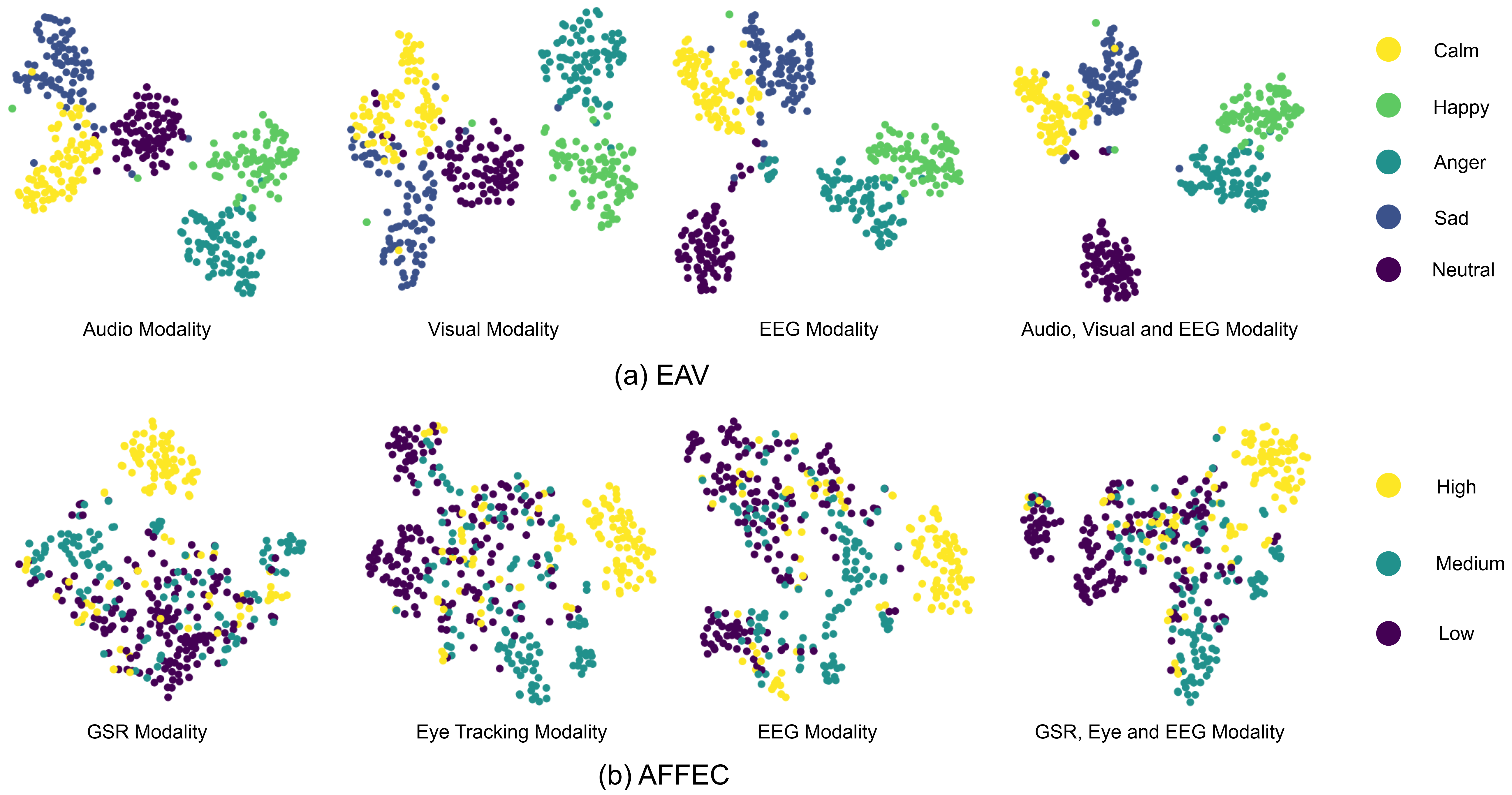}
    \caption{T-SNE visualization of learned features on the EAV and AFFEC datasets. (a) EAV dataset visualization across audio, video, EEG modalities and their fusion. (b) AFFEC dataset visualization for the Felt-Arousal task showing GSR, Eye-tracking, EEG modalities and their multi-modal fusion. Each dot represents a segment, and colors indicate emotion categories.}
    \label{fig:tSNE}
\end{figure}

\begin{figure}
    \centering
    \includegraphics[width=1\linewidth]{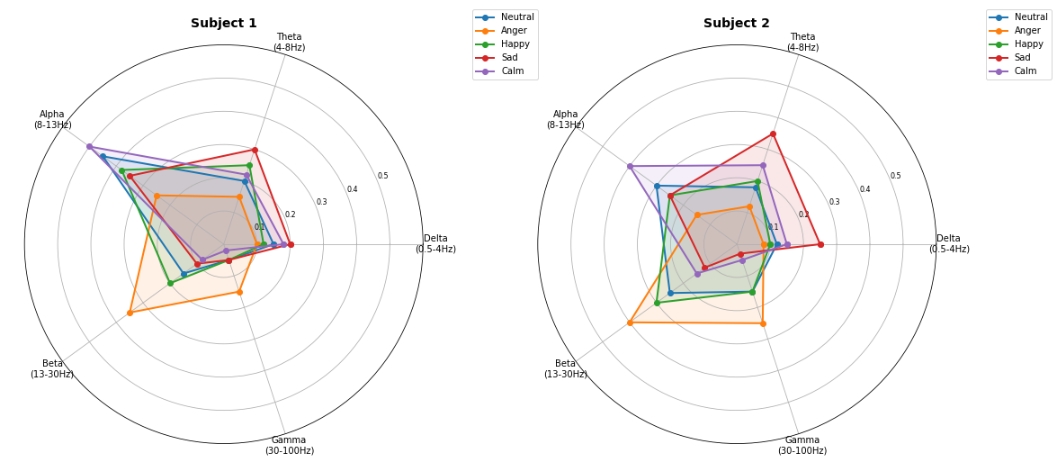}
    \caption{Inter-band attention patterns across different subjects and emotions. Subject 1 exhibits an alpha-dominant pattern while Subject 2 demonstrates a beta-gamma-dominant pattern, validating our method's ability to adapt to individual neurophysiological characteristics.}
    \label{fig:inter_band_attention}
\end{figure}

\textbf{Confusion matrix:} As shown in Figure \ref{fig:confusion_matrix},  the normalized confusion matrices with percentage values and color-bar scales clearly demonstrate per-class performance and diagonal-to-off-diagonal ratios. For the EAV dataset, diagonal accuracies range from 72.1\% (Calm) to 84.6\% (Sad), showing balanced performance across emotions rather than dominance by specific classes. Notable misclassification patterns include Happy-Calm confusion (8.0\% and 5.2\%) and Anger-Happy confusion (19.6\%), reflecting challenges in distinguishing emotions with similar arousal levels. For the AFFEC dataset, the Felt-Arousal and Felt-Valence tasks demonstrate superior performance compared to their Perceived counterparts, confirming that self-reported emotions are more reliably classified than observer-perceived emotions. The normalized percentages show Felt-Arousal achieves balanced classification (59-61\% across categories), while Felt-Valence performs better for High (64\%) and Low (68\%) than Medium (48\%). The Medium category shows increased confusion with adjacent categories, which is partly attributed to the inherent class imbalance in the AFFEC dataset where Medium samples are less represented compared to High and Low categories.

\textbf{T-SNE visualization:} Figure \ref{fig:tSNE} shows the t-SNE visualization of learned features. For the EAV dataset, the individual modality analysis reveals distinct characteristics: Audio modality demonstrates moderate clustering with reasonable emotion separation, video modality shows relatively scattered distributions with less clear boundaries due to the challenges of facial expression analysis in conversational contexts, while EEG modality presents compact and well-separated clusters, validating the reliability of physiological signals in capturing objective emotional states. For the AFFEC dataset, we focus on the best-performing Felt-Arousal task with GSR, Eye-tracking, and EEG modalities, where EEG again demonstrates superior clustering performance despite the class imbalance challenges. Most significantly, the multi-modal fusion visualization reveals dramatically enhanced feature discrimination with clearer inter-class boundaries and substantially reduced intra-class variance compared to individual modalities. This improved separability provides visual confirmation that our hypergraph-based AHFM successfully models higher-order relationships among modalities, leveraging their complementary strengths while compensating for individual limitations to achieve superior emotion recognition performance.

\textbf{Inter-band Attention Analysis:} Figure \ref{fig:inter_band_attention} visualizes the learned inter-band attention patterns from two representative subjects, revealing distinct individual patterns while maintaining neurophysiologically meaningful emotion-frequency relationships. Subject 1 exhibits an alpha-dominant response pattern with consistently elevated alpha band attention across all emotional states (peak values of 0.50 for Calm and 0.45 for Neutral), while Subject 2 demonstrates a beta-gamma-dominant pattern with pronounced high-frequency responses during high-arousal emotions (beta: 0.40, gamma: 0.25 for Anger). Despite different frequency preferences, both subjects maintain consistent emotion-arousal relationships, with high-arousal emotions showing elevated beta activity and low-arousal emotions emphasizing lower frequency bands, validating that our inter-band attention mechanism captures universal neurophysiological principles while adapting to individual characteristics.

\subsection{Cross-domain Validation: Depression Detection}

To evaluate the cross-domain generalizability of our Hyper-MML framework, we conducted additional experiments on depression detection using the MODMA dataset \cite{cai2022multi}. This validation demonstrates the framework's applicability across different mental health applications. Depression detection represents an ideal cross-domain scenario due to its high correlation with emotional disorders and significant clinical relevance.

\textbf{Dataset and Task Adaptation:} The MODMA dataset contains multi-modal data from 53 participants (24 MDD patients, 29 healthy controls) with 128-channel resting-state EEG recorded at 250Hz and audio recordings from clinical interviews. Each patient has 29 audio segments. To match sample numbers between modalities, we divided EEG data into 29 segments, yielding 638 MDD samples and 841 healthy control samples. We adapted our framework for binary classification by: (1) modifying the classifier from 5-class emotion to 2-class depression detection, (2) processing resting-state EEG through our ABEMA module, and (3) maintaining consistent audio feature extraction using openSMILE IS10 configuration. The AHFM was configured for dual-modal fusion (EEG+Audio) as video data is unavailable in MODMA.

\textbf{Baselines and Metrics:} We compared against three representative methods: MS2-GNN \cite{chen2022ms2}, a multi-modal graph neural network for depression detection with EEG-audio fusion; Effnetv2s \cite{qayyum2023high}, a CNN-based approach using EfficientNet architecture; and Emo-GCN \cite{xing2024adaptive}, an adaptive multi-graph neural network achieving state-of-the-art performance on MODMA. Performance evaluation employed four metrics: accuracy (ACC), precision (PRE), recall (REC), and F1 score.

\textbf{Results and Analysis:} Table \ref{tab:cross_domain_results} shows that Hyper-MML achieves 94.85\% accuracy, significantly outperforming MS2-GNN (86.49\%) and Effnetv2s (93.07\%), but trailing Emo-GCN (96.30\%) by 1.45\%. While both methods utilize identical EEG-Audio modalities, several factors contribute to this performance gap. Our framework was originally designed for conversational emotion recognition with task-related EEG signals, whereas depression detection relies on resting-state EEG patterns that may not fully exploit ABEMA's task-oriented frequency processing. Additionally, Emo-GCN's adaptive multi-graph architecture appears specifically optimized for depression-related neural connectivity patterns, while our hypergraph structure, though effective for complex multi-modal dynamics, may introduce unnecessary complexity for binary classification. Despite this gap, the competitive performance demonstrates effective cross-domain transfer capabilities, indicating the framework's potential for broader clinical applications through domain-specific adaptation.

\begin{table}[h]
\centering
\caption{Cross-domain validation results on MODMA depression detection dataset}
\label{tab:cross_domain_results}
\begin{tabular}{ccccc}
\hline
Method & ACC (\%) & PRE (\%) & REC (\%) & F1 (\%) \\
\hline
MS2-GNN \cite{chen2022ms2} & 86.49 & 82.35 & 87.50 & 84.85 \\
Effnetv2s \cite{qayyum2023high} & 93.07 & 92.92 & 91.76 & 93.92 \\
Emo-GCN \cite{xing2024adaptive} & \textbf{96.30} & \textbf{96.26} & \textbf{95.37} & \textbf{95.81} \\
\hline
Hyper-MML & 94.85 & 94.12 & 93.76 & 93.94 \\
\hline
\end{tabular}
\end{table}

\subsection{Potential Applications}

The practical significance of Hyper-MML lies in its potential to support real-world affective computing in clinical applications. By grounding emotion recognition in physiological EEG signals rather than relying solely on linguistic content, the framework is less dependent on explicit verbal expression, which may be particularly valuable in scenarios where communication is ambiguous or limited, such as in populations with autism spectrum disorder and depression. Moreover, the integration of EEG with audio-visual modalities through hypergraph-structured fusion enables Hyper-MML to effectively characterize inconsistencies between neural activity and external behavioral expressions, which may provide informative cues for mental health assessment and stress monitoring. In summary, the consistent robustness observed across multiple datasets suggests that our Hyper-MML model is potentially suited for practical deployment in healthcare, mental health monitoring, affect-sensitive learning, and multi-modal decision support.

\subsection{Future Work}

Several promising directions for future research toward practical application emerge. First, advanced artifact separation techniques are needed to mitigate facial muscle contamination in EEG signals during active conversation, thereby improving the reliability of physiological emotion indicators. Second, future efforts should extend the physiological signal repertoire beyond EEG to include complementary modalities (e.g., galvanic skin response, heart rate variability); develop model compression techniques that preserve recognition accuracy while enhancing computational efficiency; investigate adaptive learning mechanisms for dynamic emotional transitions in longitudinal clinical monitoring; and conduct extensive validation across diverse clinical populations to establish the framework’s diagnostic utility across mental health conditions and cultural contexts.

\section{Conclusion}\label{Conclusion and future work}

In this study, we introduced the Hypergraph Multi-Modal Learning framework (Hyper-MML) for EEG-based emotion recognition in conversations, addressing the limitations of traditional methods that primarily rely on textual information and struggle with incomplete dialogue scenarios. Our framework also specifically tackles the technical challenges inherent in multi-modal integration, including EEG's low signal-to-noise ratios, inter-subject variability, and temporal alignment issues. By integrating EEG signals with audio and video data through our novel hypergraph architecture, our framework effectively captures the intricate emotional dynamics inherent in conversational interactions while providing objective physiological indicators independent of language coherence. The proposed ABEMA encoder with hierarchical mutual-cross attention successfully models complex frequency-domain relationships in EEG signals, incorporating comprehensive artifact mitigation strategies and personalized frequency interaction patterns that adapt to individual neurophysiological characteristics. Meanwhile, the AHFM significantly enhances the model's ability to process higher-order multi-modal relationships. The proposed approach demonstrates superior performance over traditional graph-based approaches through direct modeling of triadic EEG-audio-video relationships, which leads to a more nuanced understanding of emotional states.

Our comprehensive experiments on both EAV and AFFEC datasets demonstrate that the proposed framework achieves state-of-the-art performance with substantial improvements over existing methods, maintaining robust cross-subject generalizability with only 4.25\% performance degradation in subject-independent evaluation, validating the robustness and generalizability of our hypergraph-based approach. The framework's computational efficiency, achieving rapid convergence in 12 epochs with 10.3ms inference latency, combined with successful cross-domain validation on depression detection tasks, demonstrates its practical applicability for real-world clinical deployment.

Taken together, these findings demonstrate that Hyper-MML offers an effective, efficient, and generalizable solution for EEG-centered emotion recognition in conversation, highlighting the value of hypergraph-based multi-modal learning for modeling complex cross-modal emotional dynamics.

\section*{Acknowledgments}
This work was supported by The Hong Kong Polytechnic University Start-up Fund (Project ID: P0053210), The Hong Kong Polytechnic University Faculty Reserve Fund (Project ID: P0053738), an internal grant from The Hong Kong Polytechnic University (Project ID: P0048377), The Hong Kong Polytechnic University Departmental Collaborative Research Fund (Project ID: P0056428), The Hong Kong Polytechnic University Collaborative Research with World-leading Research Groups Fund (Project ID: P0058097),  Research Grants Council Collaborative Research Fund (Project ID: C5033-24G) and in part by Shenzhen-Hong Kong Institute of Brain Science-Shenzhen Fundamental Research Institutions (Project ID: 2023SHIBS0003).

\bibliographystyle{unsrt}  
\bibliography{references}

\end{document}